\documentclass[11pt,letterpaper]{article} \usepackage[normal]{optional}
\usepackage{amsmath}
\usepackage{amssymb}
\usepackage{ifpdf}
\usepackage{algorithmic}
\usepackage{subfig}
\usepackage{wrapfig}
\usepackage{cite}

\usepackage{textcomp}

\usepackage{graphicx}
\usepackage{enumerate}

\usepackage{commands-tam}
\usepackage{numbering-tam}
\usepackage{url}
\usepackage{amsfonts}
\usepackage{amsthm}
\usepackage{fullpage}

\usepackage{times}

\def\captionfont{\small\sl}
\def\captionlabelfont{\small\bf}
{\makeatletter
 \global\let\plainfont@makecaption=\@makecaption
 \long\gdef\@makecaption#1#2{%
   \plainfont@makecaption{{\captionlabelfont #1}}{\captionfont #2}}}

\ifpdf

  \usepackage[pdftex]{epsfig}
  \usepackage[pdftex]{hyperref}

\else
    \usepackage[dvips]{epsfig}
    \newcommand{\href}[2]{#2}

\fi

{\makeatletter
 \gdef\xxxmark{%
   \expandafter\ifx\csname @mpargs\endcsname\relax 
     \expandafter\ifx\csname @captype\endcsname\relax 
       \marginpar{xxx}
     \else
       xxx 
     \fi
   \else
     xxx 
   \fi}
 \gdef\xxx{\@ifnextchar[\xxx@lab\xxx@nolab}
 \long\gdef\xxx@lab[#1]#2{\textbf{[\xxxmark #2 ---{\sc #1}]}}
 \long\gdef\xxx@nolab#1{\textbf{[\xxxmark #1]}}
}

\makeatletter
\newenvironment{floathere}[1]{\def\@captype{#1}}{}
\makeatother

\newcommand\BREAK{\textsl{BREAK}}

\begin{document}

\title{Self-Assembly of Arbitrary Shapes Using RNAse Enzymes: \\
       Meeting the Kolmogorov Bound with Small Scale Factor \\
       (extended abstract)}
\author{Erik D. Demaine\footnote{MIT Computer Science and Artificial Intelligence Laboratory,
32 Vassar St., Cambridge, MA 02139, USA, edemaine@mit.edu. This author's research was supported in part by NSF grant CDI-0941538.} \and Matthew J. Patitz\footnote{Department of Computer
Science, University of Texas--Pan American, Edinburg, TX, 78539, USA.
mpatitz@cs.panam.edu.} \and
Robert T. Schweller\footnote{Department of Computer
Science, University of Texas--Pan American, Edinburg, TX, 78539, USA.
schwellerr@cs.panam.edu.} \and Scott M. Summers\footnote{Department
of Computer Science and Software Engineering, University of Wisconsin--Platteville, Platteville, WI 53818, USA. summerss@uwplatt.edu.}}



\date{}

\maketitle
\begin{abstract}
We consider a model of algorithmic self-assembly of geometric shapes out of
square Wang tiles studied in SODA 2010, in which there are two types of
tiles (e.g., constructed out of DNA and RNA material) and one operation that
destroys all tiles of a particular type (e.g., an RNAse enzyme destroys
all RNA tiles).  We show that a single use of this destruction operation
enables much more efficient construction of arbitrary shapes.
In particular, an arbitrary shape can be constructed using an asymptotically
optimal number of distinct tile types
(related to the shape's Kolmogorov complexity),
after scaling the shape by only a logarithmic factor.
By contrast, without the destruction operation,
the best such result has a scale factor at least linear in the
size of the shape, and is connected only by a spanning tree of the scaled tiles.
We also characterize a large collection of shapes that can be constructed
efficiently without any scaling.
\end{abstract}

\setcounter{page}0
\thispagestyle{empty}
\clearpage



\section{Introduction}

DNA self-assembly research attempts to harness the power of synthetic biology to manipulate matter at the nanoscale.  The general goal of this field is to design a simple system of particles (e.g., DNA strands) that efficiently assemble into a desired macroscale object. Such technology is fundamental to the field of nanotechnology and has the potential to allow for massively parallel, bottom-up fabrication of complex nanodevices, or the implementation of a biological computer.  Motivated by experimental DNA assemblies of basic building blocks or DNA \emph{tiles}~\cite{RoPaWi04,BarSchRotWin09,CheSchGoeWin07,MaoLabReiSee00,WinLiuWenSee98,LiuShaSee99,MaoSunSee99}, the \emph{tile self-assembly model}~\cite{RotWin00} has emerged as a premier theoretical model of self-assembly.  Tile self-assembly models particles of the system by four-sided Wang tiles which float randomly in the plane and stick to one another when abutting edges have sufficient affinity for attachment.

Perhaps the most fundamental question within the tile self-assembly model is how efficiently, in terms of the number of distinct tile types needed, can a target shape be uniquely assembled.  For some special classes of shapes such as rectangles and squares, the problem has been considered in depth under a number of tile-based self-assembly models.  More generally, researchers have considered the complexity of assembling arbitrary shapes~\cite{SolWin07,DDFIRSS07,Sum09,RSAES}.  In particular, Soloveichik and Winfree~\cite{SolWin07} show that any shape, modulo scaling, can be self-assembled with a number of tile types close to the Kolmogorov complexity of the target shape.  While intriguing from a theoretical standpoint, this result has an important drawback: it assembles an arbitrarily large scaled-up version of the target shape, rather than the exact target shape.  It is conceivable that a reasonable scale factor could be tolerated in practice by simply engineering smaller tiles, but the scale factors needed for the Soloveichik-Winfree construction are unbounded in general, proportional to the running time of the Kolmogorov machine that generates the shape, which is at least linear in the size of the target shape in all cases.  This extreme resolution loss motivates the search for a practical model and construction that can achieve extremely small scale factors while retaining the Kolmogorov-efficient tile complexity for general shapes.

\paragraph{Our results.}

We achieve Kolmogorov-efficient tile complexity of general shapes with a logarithmic bounded scale factor, using the experimentally motivated \emph{Staged RNA Assembly Model (SRAM)} introduced in~\cite{SRTSARE}.  The SRAM extends the standard tile self-assembly model by distinguishing all tile types as consisting of either DNA or RNA material.  Further, in a second stage of assembly, an \emph{RNase enzyme} may be added to the system which dissolves all RNA tiles, thus potentially breaking assemblies apart and allowing for new assemblies to form.  While this modification to the model is simple and practically motivated (the idea was first mentioned in \cite{RotWin00}), we show that the achievable scale factor for Kolmogorov-efficient assembly of general shapes drops dramatically: for arbitrary shapes of size $n$, a scale factor of $O(\log n)$ is achieved, and for a large class of ``nice'' shapes, the Kolmogorov optimal tile complexity can be achieved without scaling (scale factor~$1$).  Refer to Figure~\ref{table:summary}.  Further, we show that arbitrarily large portions of infinite computable patterns of the plane can be weakly assembled within the SRAM.  Such assembly has been proved impossible in the standard tile assembly model~\cite{CCSA}, illustrating an important distinction in the power of SRAM compared to the standard tile assembly model.

In addition to tile complexity and scale factor, we also address the metrics of \emph{connectivity} and \emph{addressability}.  \emph{Full connectivity} denotes whether all adjacent tiles making up the target shape share positive strength bonds, a desirable property as it creates a stable final assembly.  All of our finite constructions are fully connected, unlike the previous result of \cite{SolWin07} which just connected a spanning tree of the scaled tiles, making for a potentially very floppy construction.  \emph{Addressability} denotes whether a construction is able to assign arbitrary binary labels to the tiles that make up the final assembly.  Addressability may have important practical applications for assemblies that are to serve as scaffolding for the fabrication of nanodevices such as circuits in which specific components must be attached to specific locations in the assembled shape.  Our $O(\log n)$-scale construction provides the flexibility to encode an arbitrary binary label within the tile types of each scaled-up position in the assembled shape, thus yielding a high degree of addressability, while our $1$-scale construction allows complete addressability.



\begin{table*}
\centering
\tabcolsep=0.5\tabcolsep
\begin{tabular}{l|c|c|c|c}
\multicolumn{1}{c|}{\textbf{General shape $S$ with $n$ points}} & \textbf{Tile Types} & \textbf{Stages} & \textbf{Scale} & \textbf{Connectivity}
\\ \hline
Previous work \cite{SolWin07} & \multicolumn{1}{|c|}{$\Theta(K(S)/\log K(S))$} & 1 & unbounded & partial
\\ \hline
Arbitrary shapes (Thm. \ref{pod_theorem_fully_connected}) & \multicolumn{1}{|c|}{$\Theta(K(S)/\log K(S))$} & 2 & $O(\log n)$ & full
\\ \hline
``Nice'' shapes (Thm. \ref{rectangle_decomp_theorem}) & \multicolumn{1}{|c|}{$\Theta(K(S)/\log K(S))$} & 2 & 1 & full
\\ \hline
\multicolumn{1}{c}{}\\

\multicolumn{1}{c|}{\textbf{Infinite computable pattern $S$}} & \textbf{Tile Types} & \textbf{Stages} & \textbf{Scale} & \textbf{Connectivity}
\\ \hline
Computable patterns (Sec. \ref{sec:computablepatterns}) & \multicolumn{1}{|c|}{$\Theta(K(S)/\log K(S))$} & 2 & 1 & partial
\\ \hline

\end{tabular}
\caption{Summary of the tile complexities, stage complexities, scale factors, and connectivity of our RNA staged assembly constructions compared with relevant previous work.  The value $K(S)$ denotes the Kolmogorov complexity of a given shape or pattern $S$, and $n$ denotes the size of (number of points in) $S$.}
\label{table:summary}
\end{table*}

\section{Preliminaries}

We work in the $2$-dimensional discrete space $\Z^2$. Let $U_2 = \{(0,1), (1,0), (0,-1), (-1,0)\}$ be the set of all unit vectors in $\mathbb{Z}^2$.
We write $[X]^2$ for the set of all $2$-element subsets of a set $X$.
All \emph{graphs} here are undirected graphs, i.e., ordered pairs $G
= (V, E)$, where $V$ is the set of \emph{vertices} and $E \subseteq
[V]^2$ is the set of \emph{edges}. A {\it grid graph} is a graph $G = (V, E)$ in which $V \subseteq \Z^2$ and
every edge $\{\vec{a}, \vec{b} \} \in E$ has the property that $\vec{a} -
\vec{b} \in U_2$.  The {\it full grid graph} on a set $V \subseteq \Z^2$ is the
graph $\fgg{V} = (V, E)$ in which $E$ contains {\it every} $\{\vec{a}, \vec{b}
\} \in [V]^2$ such that $\vec{a} - \vec{b} \in U_2$.

A \emph{shape} is a set $S \subseteq \mathbb{Z}^2$ such that $\fgg{S}$ is connected. In this paper, we consider scaled-up versions of finite shapes. Formally, if $X$ is a shape and $c \in \mathbb{N}$, then a $c$-\emph{scaling} of $S$ is defined as the set $S^c = \left\{ (x,y) \in \mathbb{Z}^2 \; \left| \; \left( \left\lfloor \frac{x}{c} \right\rfloor, \left\lfloor \frac{y}{c} \right\rfloor \right) \in X \right.\right\}$. Intuitively, $S^c$ is the shape obtained by replacing each point in $S$ with a $c \times c$ block of points. We refer to the natural number $c$ as the \emph{scaling factor} or \emph{resolution loss}. Note that scaled shapes have been studied extensively in the context of a variety of self-assembly systems \cite{SolWin07,DDFIRSS07,WinBek03,ChenGoel04,Sum09,RSAES}.

Fix some universal Turing machine $U$. The \emph{Kolmogorov complexity} of a shape $S$, denoted by $K(S)$, is the size of the smallest program $\pi$ that outputs an encoding of a list of all the points in $S$. In other words $K(S) = \min\{ |\pi| \mid U(\pi) = \langle S \rangle \}$. The reader is encouraged to consult \cite{li_vitanyi_1997} for a more detailed discussion of Kolmogorov complexity.

For more details on the specific version of the Tile Assembly Model being used, please refer to Sections \ref{sec-tam-informal}-\ref{sec-rna-removals}.

\subsection{Complexity Measures of Tile Assembly Systems}
In this paper, we are primarily concerned with measuring the ``complexity'' of a tile assembly system with respect to the following metrics.
\begin{enumerate}
    \item \textbf{Tile Complexity}. We say that the \emph{tile complexity} (sometimes called the \emph{program-size complexity} \cite{RotWin00}) is the number of unique tile types of the system.
    \item \textbf{Stage Complexity}. We say that the \emph{stage complexity} is the number of stages that a particular tile system must progress through in order to produce a terminal assembly.  (We sometimes also mention the $\BREAK$ complexity \cite{SRTSARE}, which is simply the number of $\BREAK$ stages.)
    \item \textbf{Scale Factor}. We say that a tile system produces a shape $S$ with \emph{scale factor} $c \in \mathbb{N}$ if the system uniquely produces $S^c$.
    \item \textbf{Connectivity}. When a tile system produces a terminal assembly in which not every adjacent edge interacts with positive strength, then we say that the system has \emph{partial connectivity}. On the other hand, a tile assembly system achieves \emph{full connectivity} if it only produces terminal assemblies in which every abutting edge interacts with positive strength.
    \item \textbf{Addressability}. Addressability (of the final assembly of a tile assembly system) concerns the ability of a tile system to \emph{address} or mark each tile in the final assembly with a character drawn from $\Sigma = \{0,1\}$. Note that addressability is related to the restricted notion of \emph{weak self-assembly}, formally defined in \cite{jSSADST}, and concerns the ability of tile systems to label certain (tiles placed at) locations in their final assembly as ``black'' or ``nonblack.''
\end{enumerate}

\section{The Pod Construction}

In this section, we present constructions for self-assembling slightly scaled versions of arbitrary finite shapes in the staged RNA assembly model using an asymptotically optimal number of tiles.

\subsection{Partial Connectivity Construction}

As a warmup to our main result, we obtain partial connectivity:

\begin{theorem}
\label{pod_theorem} For every finite shape $S \subset \mathbb{Z}^2$, there exists a staged RNA assembly system $\mathcal{T}_S$ that uniquely produces $S$ and moreover, $\mathcal{T}_S$ has tile complexity $O\left(\frac{K(S)}{\log K(S)}\right)$, stage complexity $2$, a scale factor of $O(\log |S|)$, and has partial connectivity.
\end{theorem}

One highlight is that the stage complexity of $\mathcal{T}_S$ is $2$, i.e., the stages in our construction consist of the initial tile addition stage followed by a single $\BREAK$ stage.  This is the fewest stages possible in any construction that makes use of the power of the RNAse enzyme. The remainder of this section is devoted to providing a proof sketch of Theorem~\ref{pod_theorem}.

\begin{figure}
\centering
\includegraphics[width=5.0in]{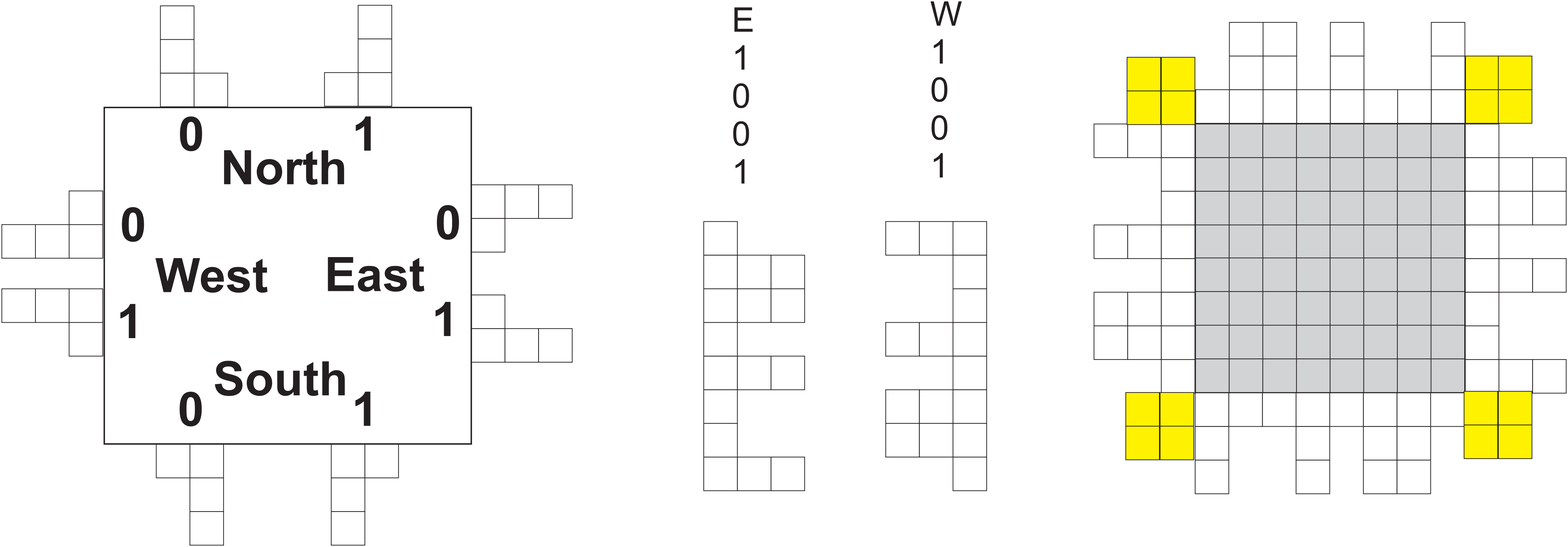} \caption{\label{fig:block-structure} \small Left: A key showing the shapes assembled for bits on each side of a block.  Middle:  An example East side and West side, each representing the bit pattern ``1001''.  Right:  An example block which has the bit pattern ``1001'' on each side.  Note that the white tiles represent the binary patterns and have null glues on their outer edges while each exposed side of each yellow block has a single strength $1$ glue exposed which is specific to its corner and direction.}
\end{figure}

\begin{figure}
\centering
\includegraphics[width=6.0in]{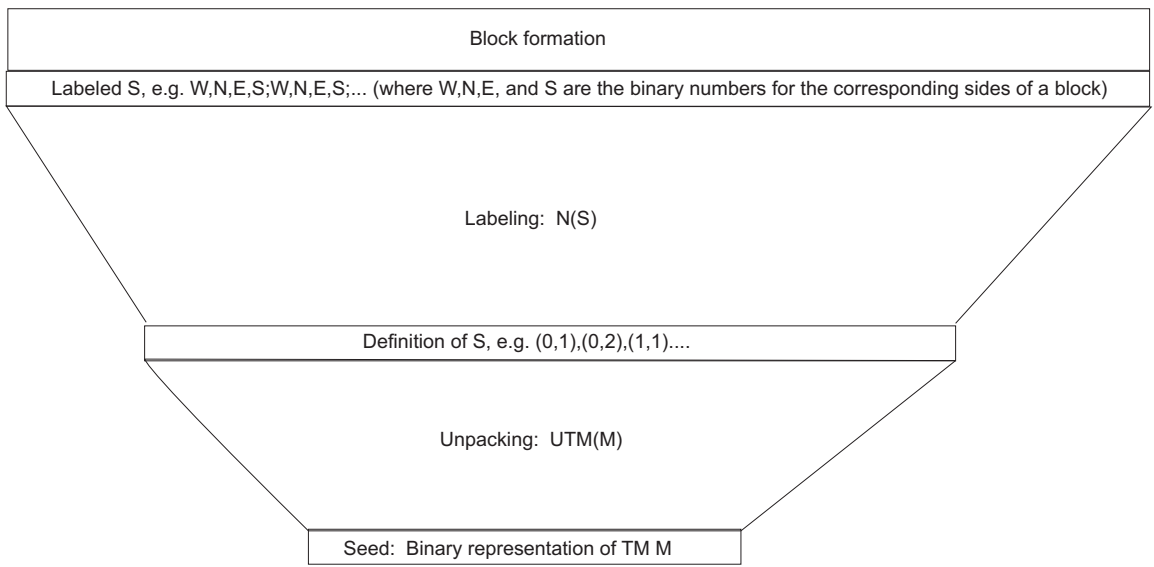} \caption{\label{fig:pod-construction-overview} \small A high level overview of the main components of the pod construction.}
\end{figure}

At a high level, the construction for Theorem~\ref{pod_theorem} works by forming a $O(\log n) \times O(\log n)$ block to represent each point in $S$.  The correct positioning of blocks is ensured by encoding binary strings that are unique to each pair of adjacent edges as ``teeth'' on the edges of the blocks.  The assembly begins with a seed, composed of RNA tile types representing a Turing machine that outputs $S$ as a list of points.  An assembly which simulates that Turing machine and then outputs definitions for each of the blocks assembles first, with all tiles being composed of RNA except for those forming the blocks, which are composed of DNA.  A $\BREAK$ operation is then performed which dissolves everything except for the DNA blocks.  These blocks then combine to form the scaled version of $S$.  Details of this construction follow. Figure~\ref{fig:block-structure} shows the basic design of the blocks used in this construction.
Figure~\ref{fig:pod-construction-overview} depicts the high level structure of this construction.
The seed row consists of a row of tiles that uniquely self-assemble into a binary representation of the shortest Turing machine $M$ that outputs the definition of a desired shape $S$ as a list of points, and then halts. Note that we use the optimal encoding scheme of \cite{SolWin07,AdChGoHu01}, which implies that the tile complexity of our construction is $O\left(\frac{K(S)}{\log K(S)}\right)$.

Assembly begins with the ``unpacking'' phase (similar to the main construction of Soloveichik and Winfree \cite{SolWin07}). Once this simulation completes, the top row of the assembly will consist of the list of points in the shape.  Next, another Turing machine, $N$, which performs the algorithm defined in Section \ref{TM-algorithm}, is simulated by the assembly.

Once $N$ halts, the top row of the assembly will consist of a sequence of binary strings that represent the binary values to be encoded along the edges of the DNA blocks. It is these blocks that will come together in a 2-handed fashion to form the final, scaled version of $S$.  The correct positioning of the blocks is ensured by the patterns of binary teeth as well as the glues on the corners of the blocks which ensure that only complementary corners of blocks can bind (e.g. the northeast corner of one block could bind only to the northwest corner of another).  Figure~\ref{fig:block-formation} gives more detail about this top row and the ``block formation'' component of the construction which actually forms the blocks.  For block edges which correspond to an outer edge of the shape $S$, instead of binary teeth a smooth edge with $0$-strength glues will be formed.  Note that the seed tiles, Turing machine simulation tiles, and white tiles from Figure~\ref{fig:block-formation} are all RNA tile types which will ultimately be dissolved by RNase enzyme in the $\BREAK$ stage.  Following the $\BREAK$, the $O(\log n) \times O(\log n)$ sized blocks representing each of the points in $S$ are free to self-assemble into the scaled up version of $S$, thus completing the construction.

\subsection{Full Addressability of Points in $S$}
\label{sect:fa1}

In the aTAM, tile types are allowed to have ``labels'' which are nonfunctional (not necessarily unique) strings associated with tile types.  Often, labels are assigned to tile types to make it easier to logically identify and group them (for instance, the ``$0$'' and ``$1$'' labels assigned to the tile types that assemble into a binary counter).  In laboratory implementations of DNA tile types, tile types are often created with the equivalent of such binary labels by the inclusion or exclusion of a hairpin loop structure which projects upward above the plane of the tile, for $0$ and $1$ respectively (a notable example of this technique is due to Papadakis, Rothemund and Winfree \cite{RoPaWi04}).  This is currently done to simplify the imaging process and therefore the detection of errors that occur in the assembly.  However, it is possible that in the future such projecting labels could be also used to create binding sites for additional materials, allowing the self-assembling structure to serve as a scaffolding for more complicated productions.  For simplicity, we assume the set of available labels to be $\Sigma = \{0,1\}$.

Here we present a construction which facilitates the arbitrary assignment of labels to subsets of locations in the final assembly.  We consider such locations to be ``addressable.''  This construction provides a method for associating labels, in the form of binary strings, with each of the points in $S$.  These binary strings will be represented by rows of tiles within the blocks, each labeled with a ``$0$'' or ``$1$.''

In the above construction, it is trivial to allow the TM $M$ encoded in the seed to also output a binary string to be used to label each/any point in $S$.  This binary string can be passed upward through the south sides of the DNA blocks so that they are represented by the labels of the tile types which form the center of each block (either in particular, designated rows or in all rows).  Of course, doing so requires an appropriate increase in tile complexity---the \emph{additional} complexity of encoding each string that will ultimately be printed on (e.g., used to address) each supertile in the final assembly. See Figure~\ref{fig:labels} for an example.

This labeling method allows bit strings of length at most equal to the width of the center portion of a DNA block (plus 2 additional tiles) to be specified for each DNA block.  Only one such unique label can be specified for each block, but the row (or rows) in which it appears can be specified by $M$.  The label can appear in any subset of the rows, or alternatively in columns. Intuitively, this is done by including a label value, which passes either upward or to the right as the center of the block assembles.  At rows (or columns) that have been specified with special markers as $M$ output the definition of the block, the label values can be ``expressed'' by tile types with the labels corresponding to the bit values.

\subsection{Full Connectivity Construction}
\label{section-full-connectivity}

Recall that for the previous constructions, the only positive strength interaction between the glues of adjacent blocks occurred at the corners of those blocks---not between the binary teeth because they have $0$-strength glues on their outer edges. We now strengthen Theorem~\ref{pod_theorem} as follows.

\begin{theorem}
\label{pod_theorem_fully_connected} For every finite shape $X \subset \mathbb{Z}^2$, there exists a staged RNA assembly system $\mathcal{T}_X$ that uniquely produces $X$ and moreover, $\mathcal{T}_X$ has tile complexity $O\left(\frac{K(X)}{\log K(X)}\right)$, $O(1)$ stage complexity, a scale factor of $O(\log |X|)$, and achieves full connectivity of the terminal assembly.
\end{theorem}

A proof sketch of Theorem~\ref{pod_theorem_fully_connected} is as follows.

In order to generate shapes with full connectivity, the scheme proposed below requires that the scaling factor be doubled from the construction of Theorem~\ref{pod_theorem} and also that, when the RNase enzyme is added, \emph{there are no remaining singleton tiles (neither DNA nor RNA) in the solution}, only the terminally produced assemblies.  The latter requirement is due to the fact that the teeth of the blocks produced have single strength glues all along their edges to which single tiles of the correct types could attach and prevent the proper connection of blocks. However, it is easy to remove this assumption by doubling the system temperature from 2 to 4, and doubling the strength of every glue that is \emph{internal} to each DNA block while maintaining single strength glues that are on the outside of the block. Note that this additional assumption is not needed for the construction for Theorem~\ref{pod_theorem} since with those blocks, there are no locations on the exposed sides to which singleton tiles could attach, only the correct and fully formed complementary blocks.

\begin{figure}
\centering
\includegraphics[width=6.0in]{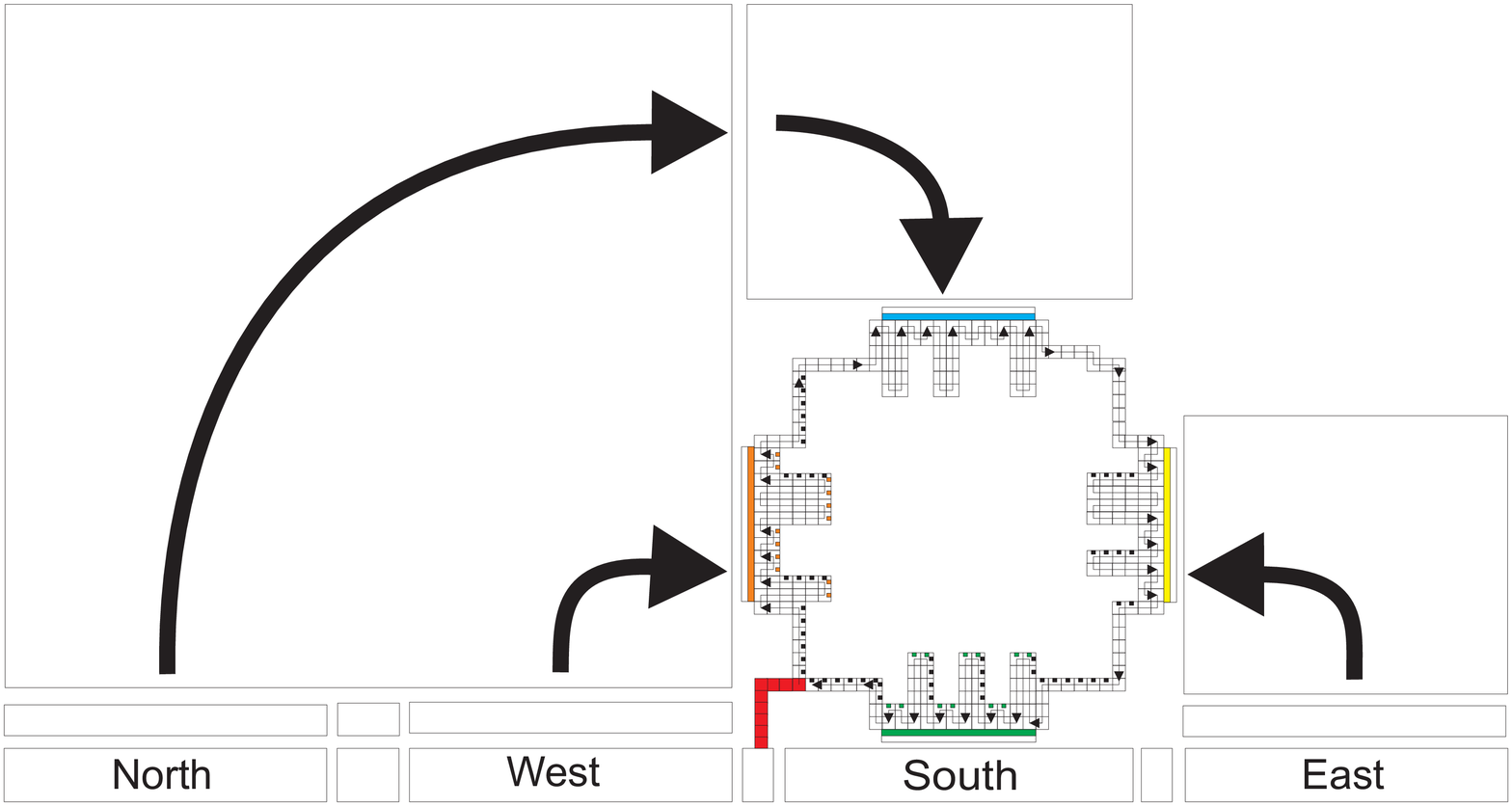} \caption{\label{fig:fully-connected-rotations} \small Positioning of block edge information.}
\end{figure}

Figure~\ref{fig:fully-connected-rotations} shows the procedure by which the values for the edges of a block are moved into the necessary positions relative to the edges of the block to be formed (analogous to what is shown for the original construction in Figure~\ref{fig:block-formation}).  It also shows how those values are turned into ``casts'' formed of RNA tiles. The high level idea is that first, before any DNA tiles can attach to the assembly, RNA tiles form a ``cast'' whose shape is the complement of the teeth of the block.  Only once the cast has completed for an edge is the assembly of the DNA teeth for that side allowed to proceed.  The order of growth for the cast is generally clockwise and is shown by arrows.  Once the cast has completed, the DNA tiles can fully form the block.  Every DNA tile has strength-$1$ glues on every edge and attaches with its south and west sides as input sides, generally forming the block from the bottom left to the top right. Details of the formation of the cast can be found in Section \ref{cast}.

Once the blocks have formed, the rest of the construction proceeds in a similar way to the previous construction.

\section{Self-Assembly of Shapes without Scaling}

Next we show how the pod construction can be modified to reduce the scale
factor from $O(\log n)$ to $1$ for a large class of finite shapes,
while still obtaining asymptotically optimal tile complexity
(according to the Kolmogorov complexity of the target shape),
using just a single $\BREAK$ stage, and
maintaining full connectivity of the final assembly.

\subsection{Self-Assembly of Rectangles of Arbitrary Dimension}
Note that the construction for Theorem~\ref{pod_theorem_fully_connected} can be modified thus giving us the following result for shapes that can be decomposed ``nicely'' into a disjoint set of constituent rectangles.

\begin{theorem}
\label{rectangle_decomp_theorem} For every finite shape $X \subset \mathbb{Z}^2$, if $X$ has a ``bounded rectangle decomposition,'' then there exists a staged RNA assembly system $\mathcal{T}_X$ that uniquely produces $X$ and moreover, $\mathcal{T}_X$ has tile complexity $O\left(\frac{K(X)}{\log K(X)}\right)$, utilizes 2 stages with a single $\BREAK$ step and achieves full connectivity of the unique terminal assembly.
\end{theorem}

We will now define ``bounded rectangle decomposition.''

\subsection{A Bounded Rectangle Decomposition of an Arbitrary Shape}
\label{section_bounded_rectangle_decomp}
\begin{figure}[t]
\centering
\begin{minipage}{0.5\textwidth}
\centering
\includegraphics[width=3.0in]{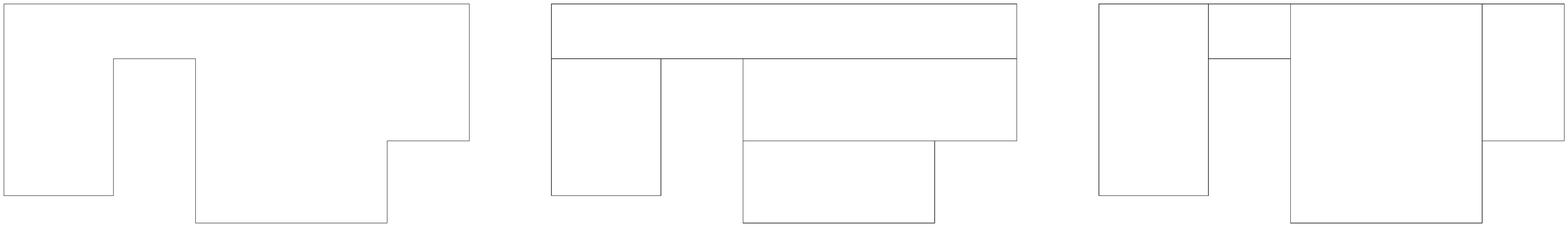} \caption{\label{fig:rectangle-decomp} \small Leftmost: a shape to be formed.  Middle and rightmost: possible rectangle decompositions of the shape.}
\end{minipage}%
\hfill
\begin{minipage}{0.48\textwidth}
\centering
\includegraphics[width=2.0in]{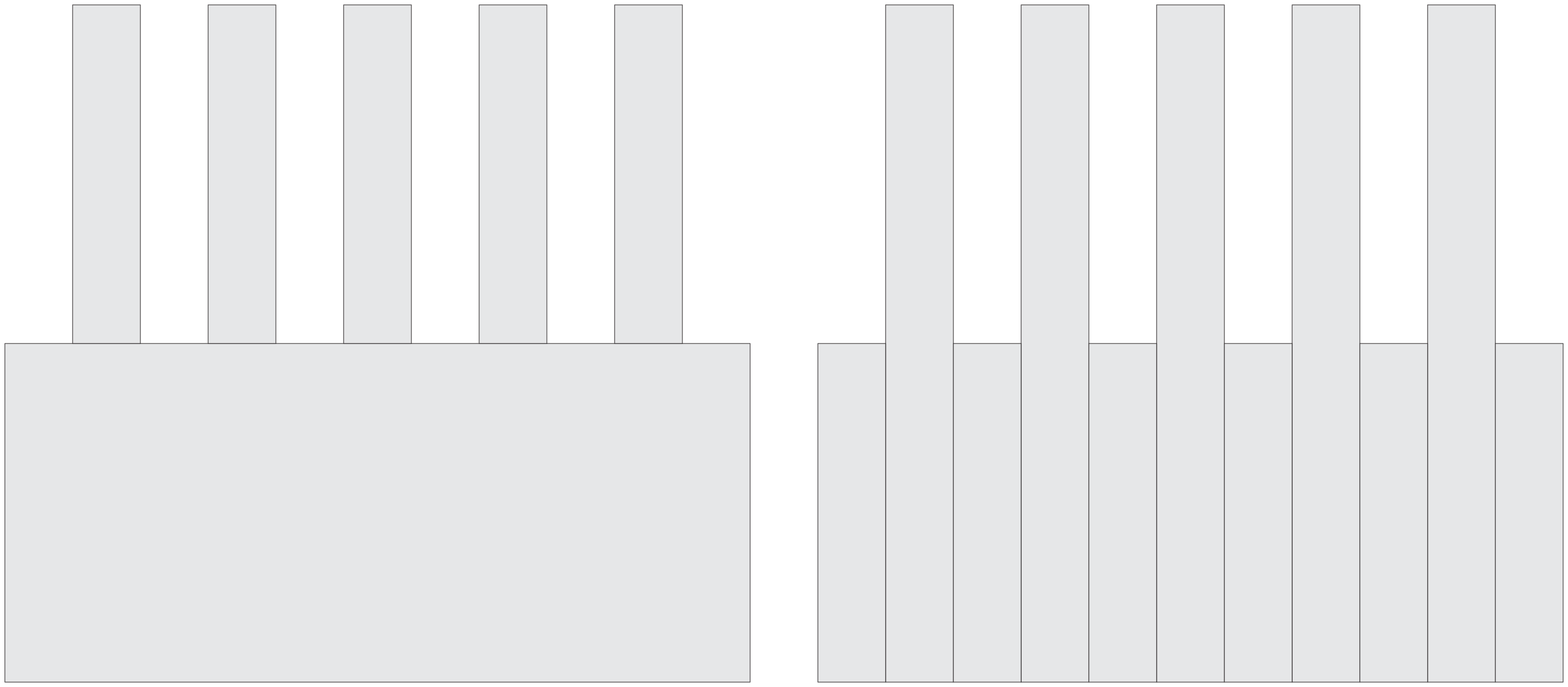}\caption{\label{fig:shape-divisions} \small Possible decomposition of a shape into rectangles. Note that the decomposition on the right is ``nicer'' (or perhaps more efficient) than the decomposition on the left since the former involves longer---and therefore, potentially more unique---interfaces. }
\end{minipage}
\end{figure}

The leftmost image in Figure~\ref{fig:rectangle-decomp} shows an example of a simple target shape to be assembled. The middle and rightmost images show two different possible rectangle decompositions of that shape.  Instead of having binary teeth along the full edges of each constituent rectangle, binary teeth need only be present at the locations where rectangles must come together, i.e., at the \emph{interface} between two rectangles. The remainder of the outside edges can be made smooth, with $0$-strength glues. Throughout this section, $X$ denotes an arbitrary finite shape.

A shape $R$ is a rectangle if $R = \{ (x,y) \in \mathbb{Z}^2 \mid a  \leq x < m+a \textmd{ and } b \leq y < n+b \textmd{ for some } a,b,m,n\in\mathbb{N} \}$. In this case, we say that $R$ is a rectangle of width $m$ and height $n$ positioned at $(a,b)$. We say that $\mathcal{R}(X) = \{R_i\}_{i=0}^{k}$, for some $k \in \mathbb{N}$ is a \emph{rectangle decomposition} of $X$ if for all $0 \leq i < k$, $R_i$ is a non-empty rectangle, $\bigcup_{i=0}^{k-1}{R_i} = X$ and for all $i,j \in \mathbb{N}$ such that $i\ne j$, $R_i \cap R_j = \emptyset$. See Figure~\ref{fig:rectangle-decomp} for examples.
Let $\mathcal{R} = \{ R_i \}_{i=0}^{k-1}$ be a rectangle decomposition of $X$ and suppose that $R_i$ and $R_j$ are rectangles in $\mathcal{R}$. For each $\vec{u} \in U_2 = \{(0,1),(1,0),(0,-1),(-1,0)\}$, denote as $I^{\vec{u}}(R_i,R_j)$ the \emph{interface between rectangles} $R_i$ \emph{and} $R_j$ \emph{in direction} $\vec{u}$, i.e., $I^{\vec{u}}(R_i,R_j)$ is the set of all points $(x,y) \in R_j$ such that $(x,y) = (w,z) + \vec{u}$ for some $(w,z) \in R_i$. It is easy to see that, for any rectangle decomposition $\mathcal{R}$, $I^{\vec{u}}(R_i,R_j)$ is the unique interface in direction $\vec{u}$ between $R_i$ and $R_j$ \emph{or} $I^{\vec{u}}(R_i,R_j) = \emptyset$. For each $\vec{u} \in U_2$, the \emph{length} of an interface $I^{\vec{u}}(R_i,R_j)$ is $|I(R_i,R_j)|$. For each $\vec{u} \in U_2$, we say that the \emph{orientation of an interface} $I^{\vec{u}}(R_i,R_j)$ is \emph{horizontal} if $\vec{u} \in \{(1,0),(-1,0)\}$ and \emph{vertical} if $\vec{u} \in \{(0,1),(0,-1)\}$. We say that $R_i$ and $R_j$ are \emph{adjacent} if $I^{\vec{u}}(R_i,R_j) \ne \emptyset$ for some $\vec{u} \in U_2$.

\begin{definition}
\label{bounded}
Let $\mathcal{R} = \{R_i\}_{i=0}^{k-1}$ be a rectangle decomposition of $X$. We say that $\mathcal{R}$ is a \emph{bounded rectangle decomposition} if:
\begin{enumerate}
    \item for each $l \in \mathbb{N}$, $\left|\left\{\left.\left|I^{\vec{u}}(R_i,R_j)\right| = l \; \right| \; \vec{u} \in U_2, i,j \in \mathbb{N} \textmd{ and } R_i,R_j \in \mathcal{R} \right\}\right| \leq 2^{\left\lfloor \frac{l-12}{4} \right\rfloor}$ and
    \item for all $R_i,R_j \in \mathcal{R}$, if $R_i$ and $R_j$ are adjacent (in some particular direction $\vec{u} \in U_2$), then $\left|I^{\vec{u}}(R_i,R_j)\right| \geq 16$.
\end{enumerate}
\end{definition}
Definition~\ref{bounded} is motivated by the way we will ultimately construct self-assembly interfaces between DNA supertiles in our forth-coming construction (discussed in the next section): each supertile-supertile interface of length $l$ can play host to at most $\left\lfloor \frac{l - 12}{4}\right\rfloor$ binary ``teeth'' since we will use $6$ tiles for each corner piece and $4$ tiles for the representation of each bit in the interface.

Intuitively, the first condition in Definition~\ref{bounded} says that there cannot be ``too many'' (i.e., roughly exponentially-many) interfaces of each length in $\mathcal{R}$, whereas the second condition is merely saying that every non-empty interface must be at least a certain length.

\subsection{Full Addressability of Every Tile in the Final Assembly}

\begin{figure}
\centering
\includegraphics[width=4.0in]{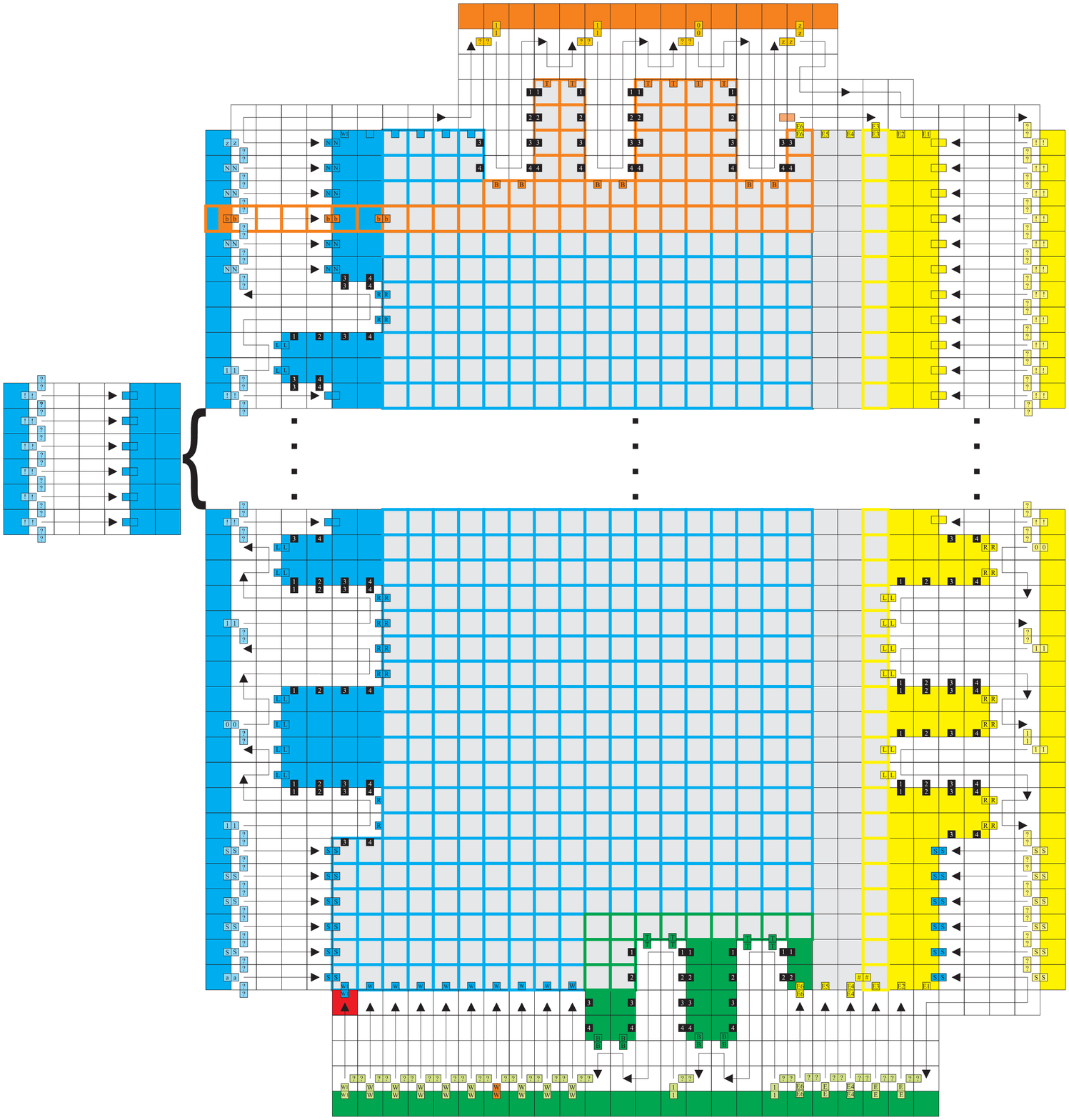} \caption{\label{fig:fully-addressed-and-connected-rect} \small Schematic of the self-assembly of a fixed-width, fully addressable rectangle that will ultimately (after the RNAse enzyme is applied) participate in the self-assembly of a fully connected and fully addressable unique terminal assembly (see Section~\ref{cast} for details about the formation of the RNA cast).  The cast forms as a single path around the entire perimeter, beginning at the bottom left side.  Shaded/colored tiles are DNA tiles while white tiles are RNA.  Only colored, non-grey tiles are allowed to assemble before the entire cast assembles. We depict single strength bonds as little colored (and labeled) squares along the edges of tiles. Arrows represent double strength bonds between contiguous groups of tiles through which they pass. Note that many (many!) glue labels are omitted in this version of the paper for the sake of clarity (and sanity!).}
\end{figure}

In this section, we sketch a construction utilizing a single $\BREAK$ step which assembles shapes (that can be ``nicely'' decomposed into rectangles) with no scaling, full connectivity, and full addressability (in the form of specifying either a $0$ or $1$ label to appear in every single tile position of the final assembly). This strengthens Theorem~\ref{rectangle_decomp_theorem} with respect to addressability but with an additional increase in tile complexity of $O(K(B))$, where $B \subseteq X$ is the set of points to be addressed, i.e., the set of points in the final assembly at which tiles labeled with a ``1'' are placed, as well as requiring an additional constraint on the rectangles contained within the rectangle decomposition.  For this construction, we require that there is some constant $k \in \mathbb{Z}^+$ that bounds at least one dimension of every rectangle in every valid rectangle decomposition. In other words, every rectangle, although potentially arbitrarily long (or wide) in one dimension, must be no longer or wider \emph{in the other dimension}, than $k$ tiles.

\begin{figure}[htp]
\centering
    \subfloat[][An example target shape $X$ and a candidate bounded-rectangle decomposition.]{%
        \label{fig:example_shape_with_decomp}%
        \includegraphics[width=1.8in]{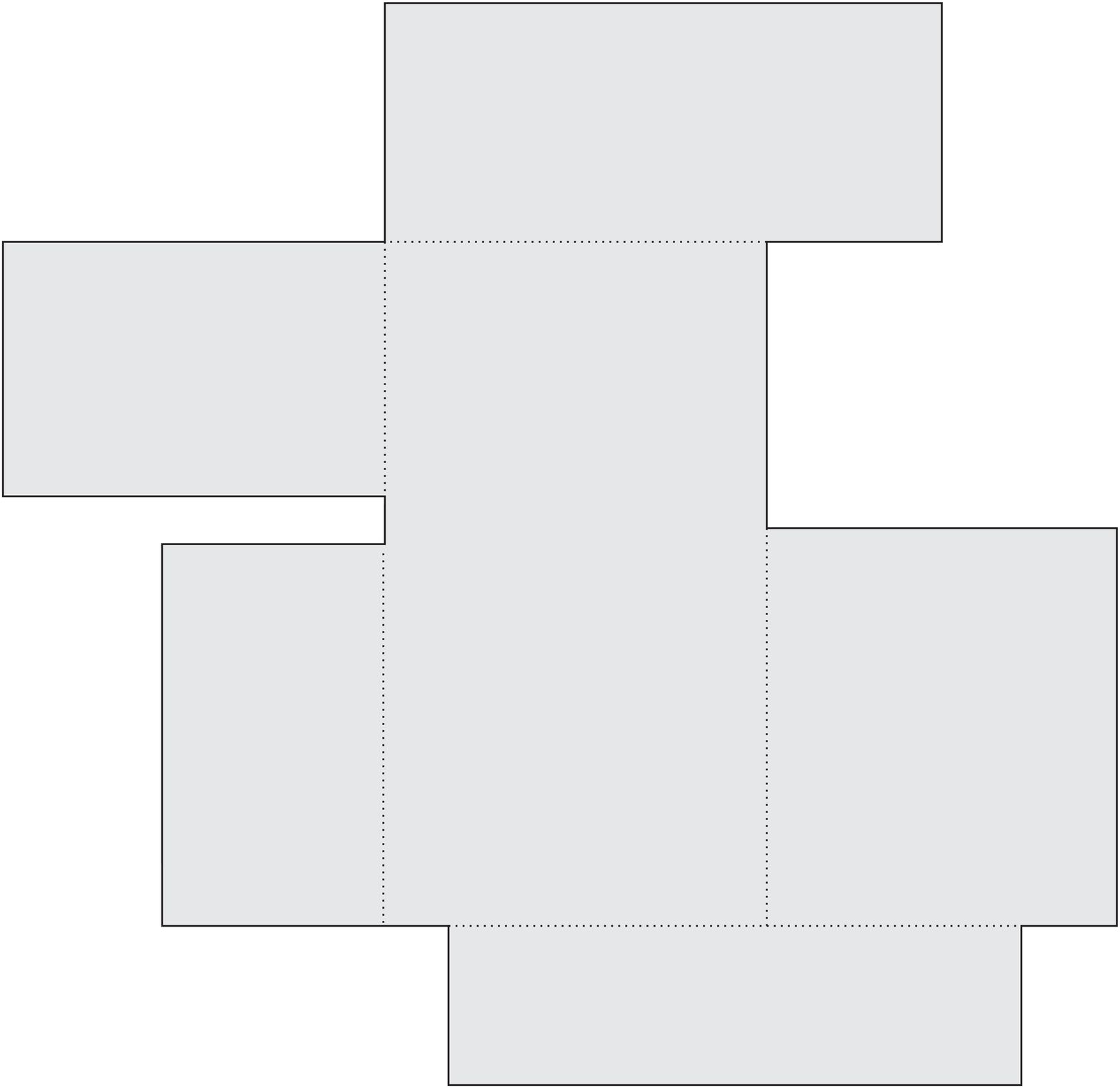}}%
        \hfil\hfil
    \subfloat[][The corresponding supertiles. For the sake of example, assume the width of each supertile must not exceed that of the bottommost supertile.]{%
        \label{fig:example_shape_supertiles}%
        \hspace{1em}%
        \includegraphics[width=1.8in]{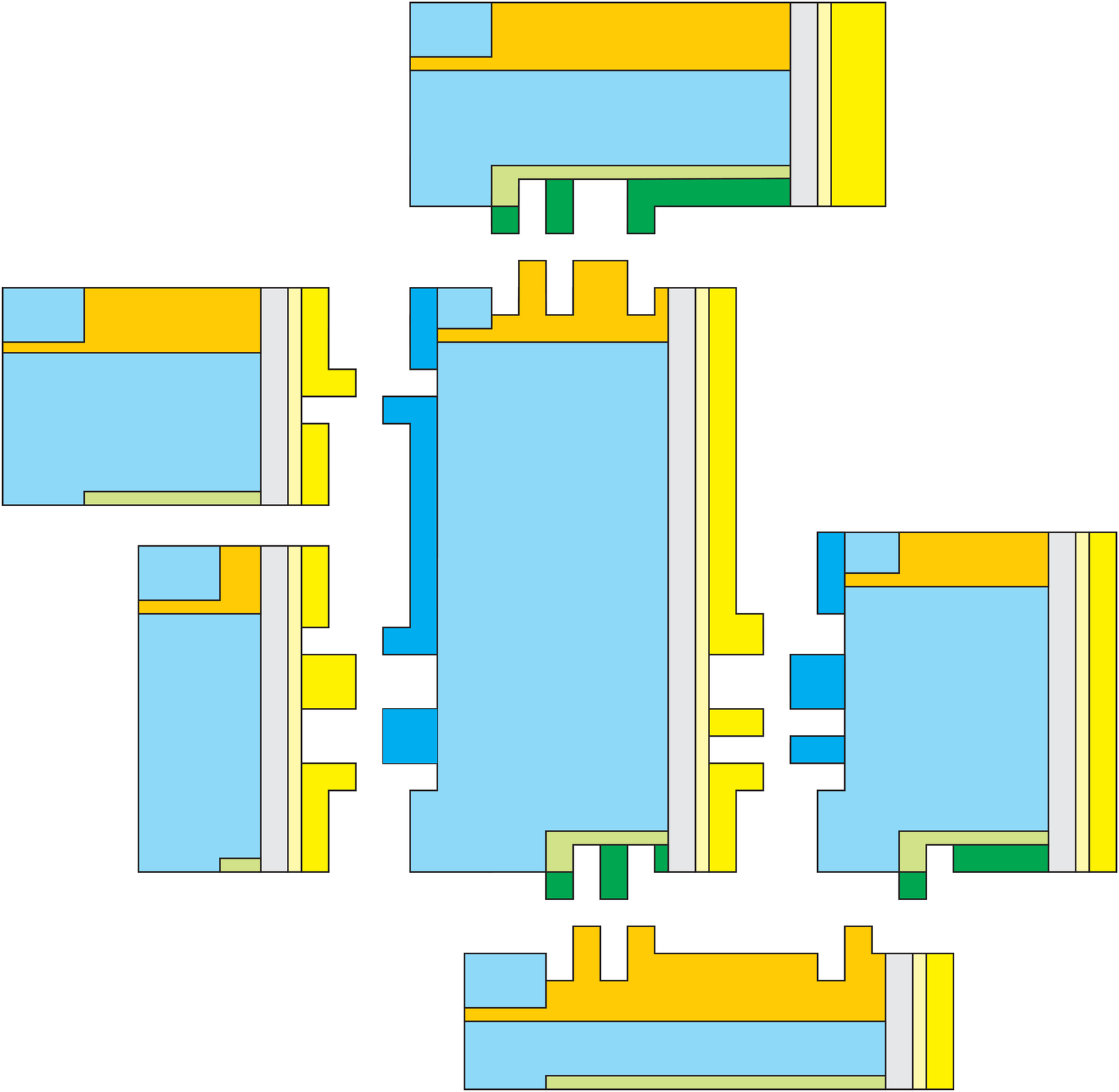}
        \hspace{1em}}
        \hfil\hfil
    \subfloat[][The final assembly.]{%
        \label{fig:example_shape_final}%
        \includegraphics[width=1.8in]{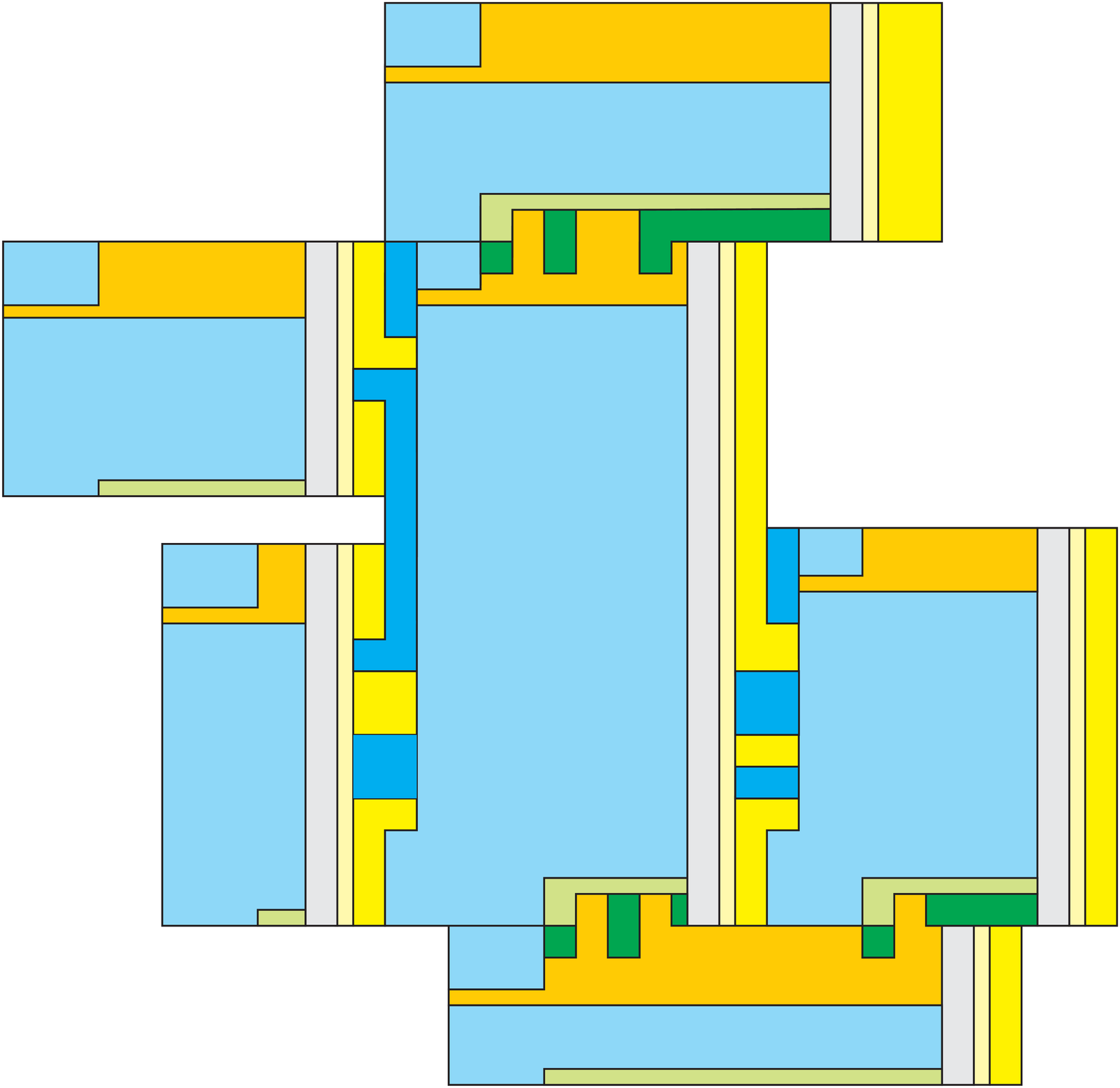}}
    \caption{\small \label{fig:example_shape} High-level schematic of the full addressability construction. Each individual supertile is colored so as to correspond to the more detailed Figure~\ref{fig:fully-addressed-and-connected-rect}. Note that the bottommost supertile attaches via two north-facing interfaces!}
\end{figure}

The details of how the rectangular blocks for this construction are formed are depicted in Figure~\ref{fig:fully-addressed-and-connected-rect}.  Our construction can be thought of to proceed in four logical phases: the unpacking process, self-assembly of the RNA cast, self-assembly of the rectangular supertiles, and self-assembly of the target shape.  The main difference with the previous construction is in the complexity of the cast and the order of assembly of the tiles forming the rectangular supertiles.  At a high level, this is due to the fact that information about the specific labels, and therefore tile types, which need to eventually occupy every single position must be propagated from the casts into the forming rectangular supertiles.  It is this need which forces the constraint on one dimension of each rectangle, and the fact that the construction retains full connectivity forces the positioning of the glues on the cast which propagate the information to be greatly complicated.  Details of this construction can be found in Section \ref{sec-full-addr-details}, and a high level schematic can bee seen in Figure \ref{fig:example_shape}.

\subsection{Weak Self-Assembly of Computable Patterns}
\label{sec:computablepatterns}
\emph{Weak self-assembly} is a general notion of self-assembly that applies to the self-assembly of patterns that are in some sense ``painted'' on a canvas of tiles that strictly contains $X$ (as opposed to \emph{strict self-assembly}, which pertains to the self-assembly of a given target shape and nothing else). Intuitively, we say that a pattern $X \subseteq \mathbb{Z}^2$ weakly self-assembles if there is a tile system that places special ``black'' marker tiles on---and only on---every point that belongs to the set $X$.

Our final construction self-assembles an arbitrarily ``large'' (square) portion of any computable pattern,
with the size of the portion of the pattern being determined simply by how long the self-assembly is allowed to proceed before the $\BREAK$ operation is performed.  This clearly demonstrates the fact that staged self-assembly with DNA removals is strictly more powerful than the standard aTAM, in terms of the weak self-assembly of patterns, as it was shown in \cite{CCSA} that there are (decidable) patterns which cannot weakly self-assemble in the aTAM.

\begin{figure}
\centering
\includegraphics[width=2in]{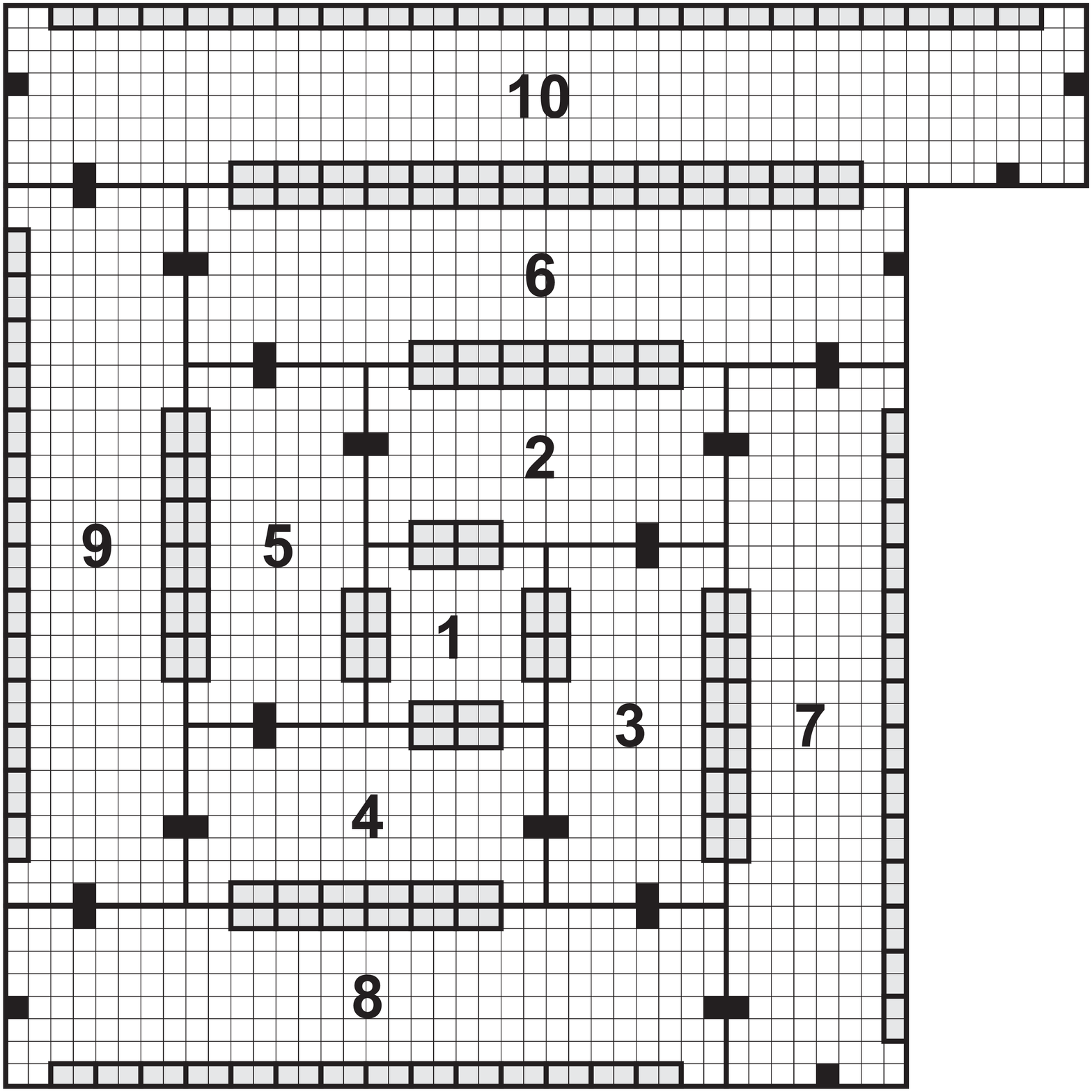} \caption{\label{fig:patterns} \small Any computable pattern in $\mathbb{Z}^2$ can be weakly self-assembled.  Note that if the pattern is infinite, then an arbitrarily large portion of it can weakly self-assemble (or the entire infinite shape if the first stage is allowed to terminate in its infinite assembly).  Essentially, an assembly that simulates a Turing machine with RNA tiles and creates pods for DNA tile rectangles with constant width (or height) and increasingly large height (or width) is first created, each with the tile labels (or colors) corresponding to the correct portion of the pattern.  Then, after the $\BREAK$ operation, the rectangles assemble to fill the plane.  This figure shows the first $10$ rectangles that are assembled.  The dark grey portions represent the locations of the binary teeth that ensure correct assembly.  Notice that the size of those portions, and therefore the number of binary teeth, steadily increases.  This forms a fully addressable assembly that is not fully connected since the tiles that do not belong to the binary teeth of each rectangle or contain black squares (which represent additional single strength bonds) have 0-strength glues.}
\end{figure}

The algorithm directing this self-assembly essentially creates larger and larger, fully addressable (although not fully connected) rectangles that ultimately come together in a ``spiral'' fashion.  Figure~\ref{fig:patterns} demonstrates the basic idea for the formation of labeled rectangles of increasing size that will ultimately combine to weakly self-assemble arbitrary computable patterns.  The darker grey portions represent the binary teeth used to connect the rectangles. Note that these rectangle-rectangle interfaces get larger as the rectangles grow out from the center.  For infinite patterns, the portion of the construction that performs the Turing machine computation and outputs the definitions of the rectangles must be slightly modified so that the rectangles are formed on the left side of north-growing simulation, enumerated one after another.  This allows for an arbitrarily large portion of such a pattern to be weakly self-assembled by simply allowing the assembly to proceed for a ``long enough'' period of time before performing the $\BREAK$ operation.

\bibliographystyle{amsplain}
\bibliography{tam,main}

\clearpage
\appendix
\section{Appendix}

\subsection{Informal Description of the Two-Handed Abstract Tile Assembly Model}
\label{sec-tam-informal}

In this section we give a brief informal sketch of a variant of Erik Winfree's abstract Tile Assembly Model (aTAM) \cite{Winf98,Winfree98simulationsof} known as the \emph{two-handed} aTAM, which has been studied previously under various names \cite{AGKS05,DDFIRSS07,Winfree06,Luhrs08,AdlCheGoeHuaWas01,Adl00}.  Please see \cite{SFTSAFT} for a more detailed description of the model and our notation.

A \emph{tile type} is a unit square with four sides, each having a \emph{glue} consisting of a \emph{label} (a finite string) and \emph{strength} (0, 1, or 2).
We assume a finite set $T$ of tile types, but an infinite number of copies of each tile type, each copy referred to as a \emph{tile}. A \emph{supertile} (a.k.a., \emph{assembly}) is a positioning of tiles on the integer lattice $\Z^2$.
Two adjacent tiles in a supertile \emph{interact} if the glues on their abutting sides are equal.
Each supertile induces a \emph{binding graph}, a grid graph whose vertices are tiles, with an edge between two tiles if they interact.
The supertile is \emph{$\tau$-stable} if every cut of its binding graph has strength at least $\tau$, where the weight of an edge is the strength of the glue it represents.
That is, the supertile is stable if at least energy $\tau$ is required to separate the supertile into two parts.
A \emph{tile assembly system} (TAS) is a pair $\calT = (T,\tau)$, where $T$ is a finite tile set and $\tau$ is the \emph{temperature}, usually 1 or 2.
Given a TAS $\calT=(T,\tau)$, a supertile is \emph{producible} if either it is a single tile from $T$, or it is the $\tau$-stable result of translating two producible assemblies.
A supertile $\alpha$ is \emph{terminal} if for every producible supertile $\beta$, $\alpha$ and $\beta$ cannot be $\tau$-stably attached.
A TAS is \emph{directed} (a.k.a., \emph{deterministic}, \emph{confluent}) if it has only one terminal, producible supertile.
Given a connected shape $X \subseteq \Z^2$,  a TAS $\calT$ \emph{produces $X$ uniquely} if every producible, terminal supertile places tiles only on positions in $X$ (appropriately translated if necessary).

\subsection{RNA tiles and RNAse enzyme}

In this paper, we assume that each tile type is defined as being composed of either DNA or RNA.  By careful selection of the actual nucleotides used to create the glues, tile types of any combination of compositions can bind together. The utility of distinguishing RNA-based tile types comes from that fact that, at prescribed points during the assembly process, the experimenter can add an RNAse enzyme to the solution which causes all tiles composed of RNA to dissolve.  We assume that, when this occurs, all portions of all RNA tiles are completely dissolved, including glue portions that may be bound to DNA tiles, returning the previously bound edges of those DNA tiles to unbound states.

More formally, for a given supertile $\Gamma$ that is stable at temperature $\tau$, when the RNAse enzyme is added, all positions in $\Gamma$ which are occupied by RNA tiles change to the empty tile.  The resultant supertile may not be $\tau$-stable and thus defines a multiset of subsupertiles consisting of the maximal stable supertiles of $\Gamma$ at temperature $\tau$, denoted by $\BREAK_\tau(\Gamma)$.

The plausibility of this model was mentioned already by Rothemund and Winfree
in 2000 \cite{RotWin00},
but it was not formalized into a model until SODA 2010 \cite{SRTSARE}
when it was combined with the idea of staged assembly \cite{DDFIRSS07}.

\subsection{Staged assembly with RNA removals}
\label{sec-rna-removals}
Staged assembly consists of a finite sequence of stages, modeling the actions taken by an experimenter (e.g., bioengineer). A stage assembly system specifies each stage as either a tile addition stage, in which new tile types are added to the system, or an enzyme stage, in which assembled supertiles are broken into pieces by deleting all occurrences of RNA tile types. In both cases, each stage consists of an initial set of preassembled supertiles from the previous stage, unioned with a new set of tile types in the case of a tile addition stage, or the current supertile set broken into subsupertiles (which may then be able to bind to each other) in the case of an enzyme stage. From this initial set, the output of the stage is determined by the
two-handed assembly model, and the stage ends once all supertiles are terminal, meaning that no further bindings can occur.  It is only at this point which the next stage can be initiated.

\subsection{Algorithm for Turing Machine $N$ of Theorem \ref{pod_theorem}}
\label{TM-algorithm}

\begin{center}
\begin{minipage}{5.5in}
\begin{algorithmic}
\STATE $c \leftarrow 0$
\FORALL{$p \in S$}
\FORALL{$d \in \{North,East,South,West\}$}
\IF{side $d$ of $p$ has not been visited}
\IF{there exists $q \in S$ and $q$ borders $p$ in direction $d$}
\STATE assign value $c$ to direction $d$ of $p$ and mark that side \emph{visited} \\
\STATE assign value $c$ to direction $-d$ of $q$ and mark that side \emph{visited} \\
\STATE $c \leftarrow c+1$ \\
\ELSE
\STATE assign the $null$ value to direction $d$ of $p$
\ENDIF
\ENDIF
\ENDFOR
\ENDFOR
\STATE $l \leftarrow \lceil 1+\lg c \rceil$ \\
\FORALL{$p \in \textbf{S}$}
\STATE Write the binary numbers assigned to each side of $p$, padded to length $l$, separated by $2$ blank symbols and in the order $West$, $North$, $East$, $South$ \\
\ENDFOR
\end{algorithmic}
\end{minipage}
\end{center}

\subsection{Details of Block Formation for Theorem \ref{pod_theorem}}

\begin{floathere}{figure}
\centering
\includegraphics[width=5.5in]{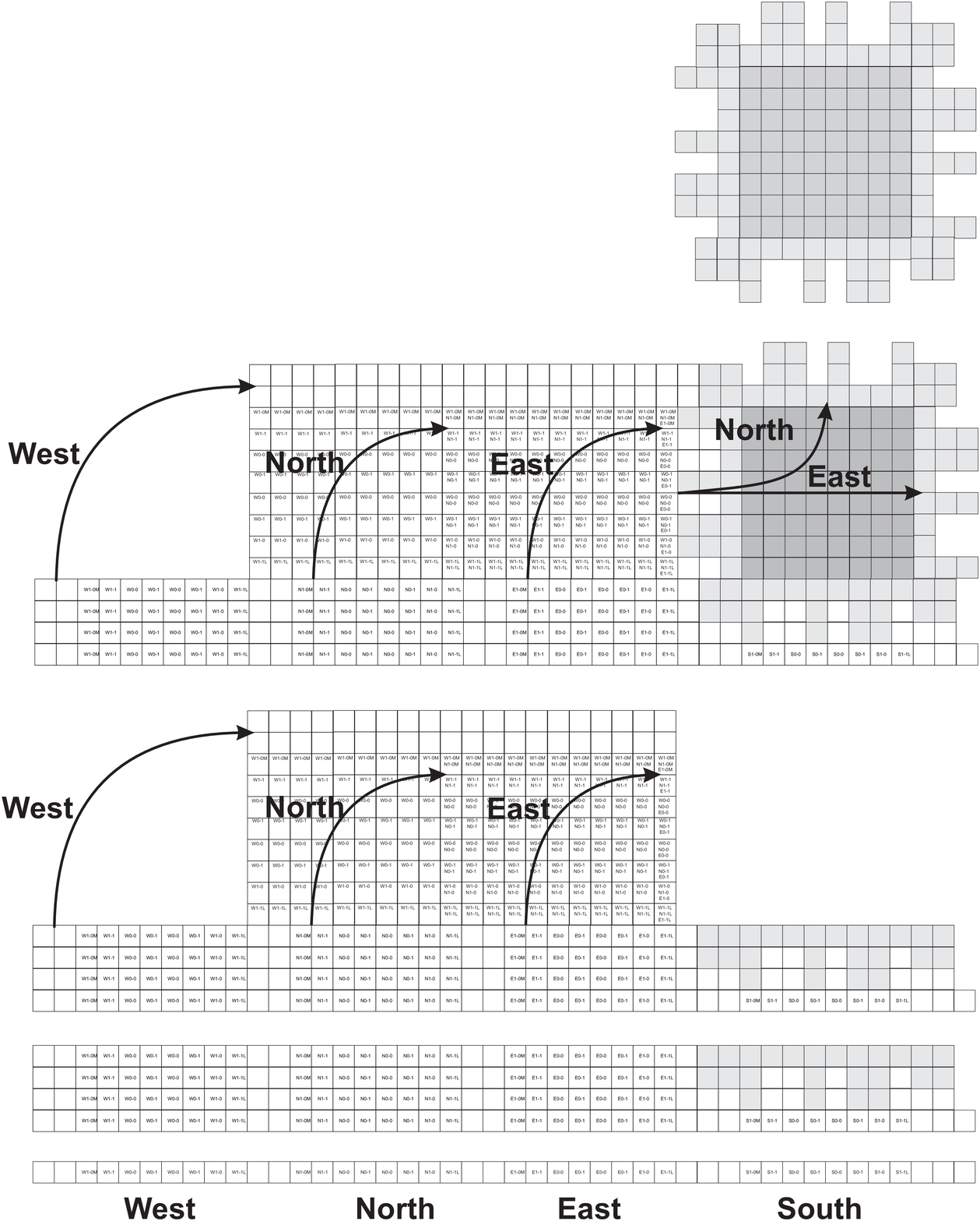} \caption{\label{fig:block-formation} \small Note that the bottom row in this figure represents a portion of the ``labeled S,'' namely a section corresponding to the definition of a single block with the binary numbers for the West, North, East, and South sides arranged from left to right.  RNA tiles are white and DNA tiles are shaded. Initially, there is a single row of tiles which represents the block.  Then, three rows grow upward which assemble the South side of the block and propagate the values for the West, North, and East sides upward. Next, the values for the West, North, and East sides are rotated up and combined together at the West side of the block.  At this point, the West side forms while propagating the information for the East and North sides. The information for the North side is rotated upward, and then the North and East sides form.  Finally, the {R}{N}ase enzyme is added and the RNA tiles are dissolved.}
\end{floathere}

\subsection{Addressability by Binary Strings}

\begin{floathere}{figure}
\centering
\includegraphics[width=3.0in]{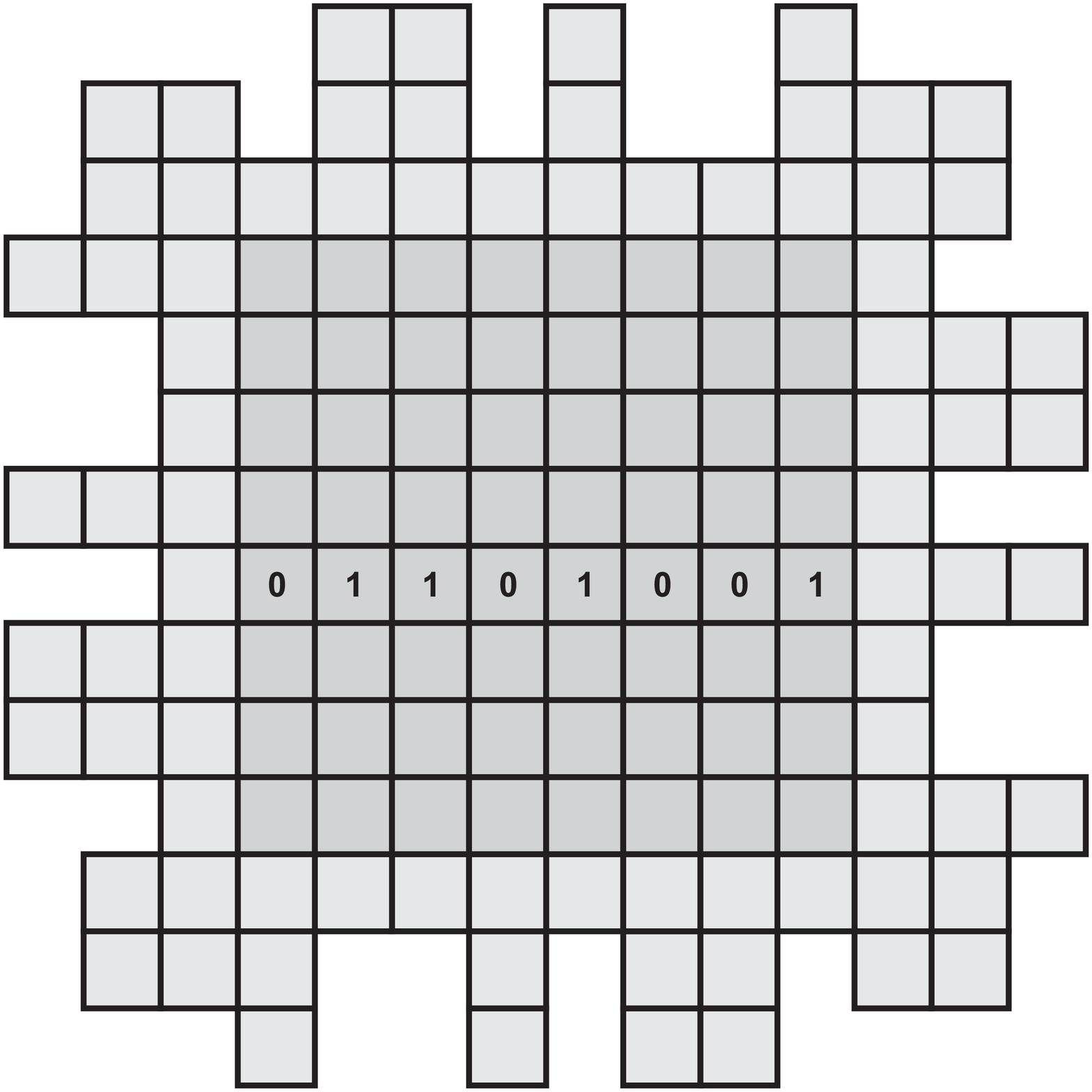}
\hspace{20pt}
\includegraphics[width=3.0in]{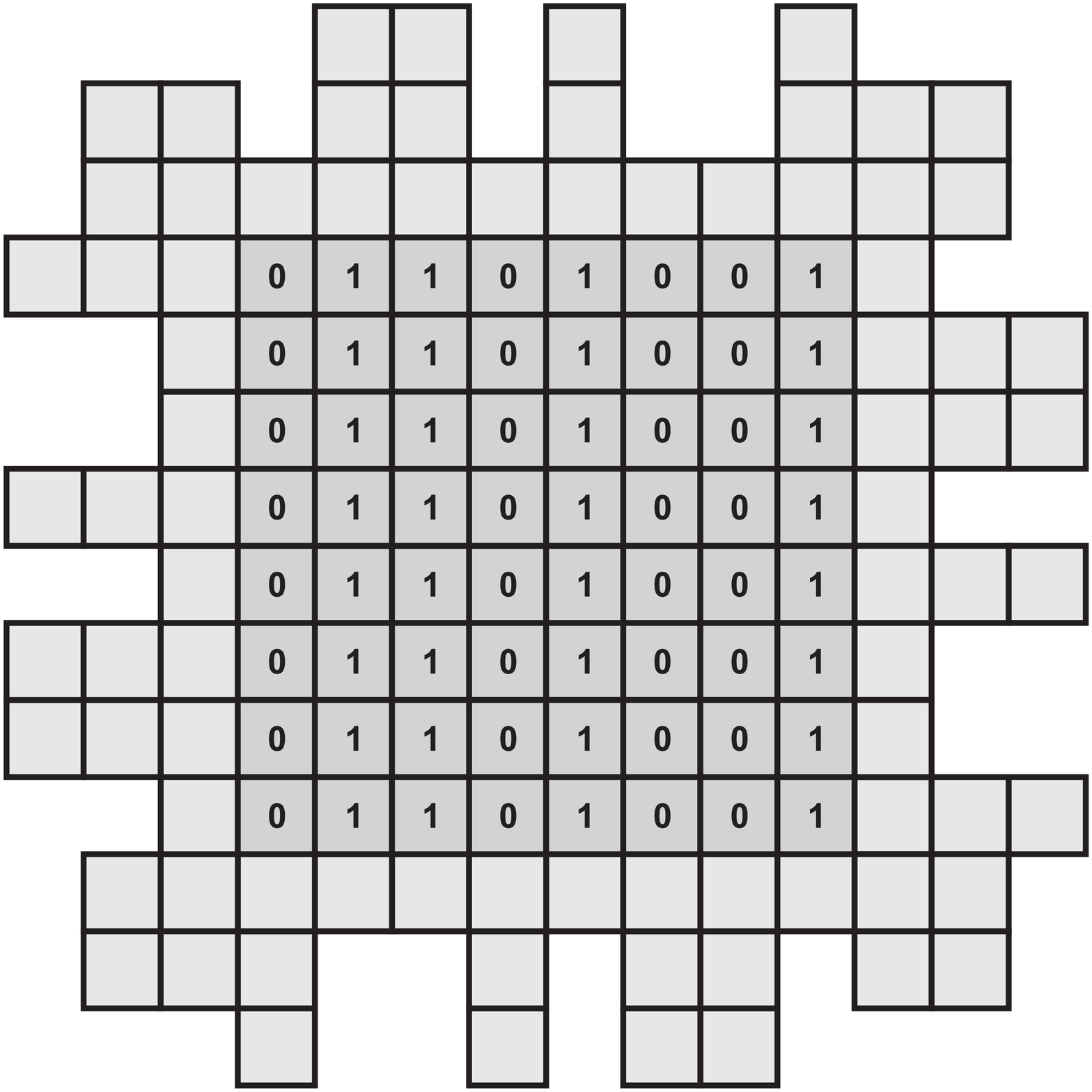}
\caption{Example blocks which are labeled with the binary string ``$01101001$.''  The block on the left has the label in only one row, while the block on the right has it in all rows.}
\label{fig:labels}
\end{floathere}

\subsection{Details of the Cast Formation for Theorem \ref{pod_theorem_fully_connected}}
\label{cast}
Without loss of generality, we can discuss the formation of the casts as forming in distinct and well-ordered steps since cooperation between tile types is utilized to ensure that each modular component forms in the correct order.  Figure \ref{fig:fully-connected-rotations} shows how the information about the bit patterns for the edges is moved into the appropriate locations.  First, the bottom row encoding the edge information forms.  Next, as depicted by the large black arrows, blocks of tiles assemble which rotate that information into position, stopping once the colored (green, orange, blue, and yellow) rows of tiles are assembled.  At this point, all of the information needed to assemble a particular block is positioned to allow the actual cast, with the correctly shaped bumps and dents, to form.

\begin{figure}
\centering
\includegraphics[width=4.0in]{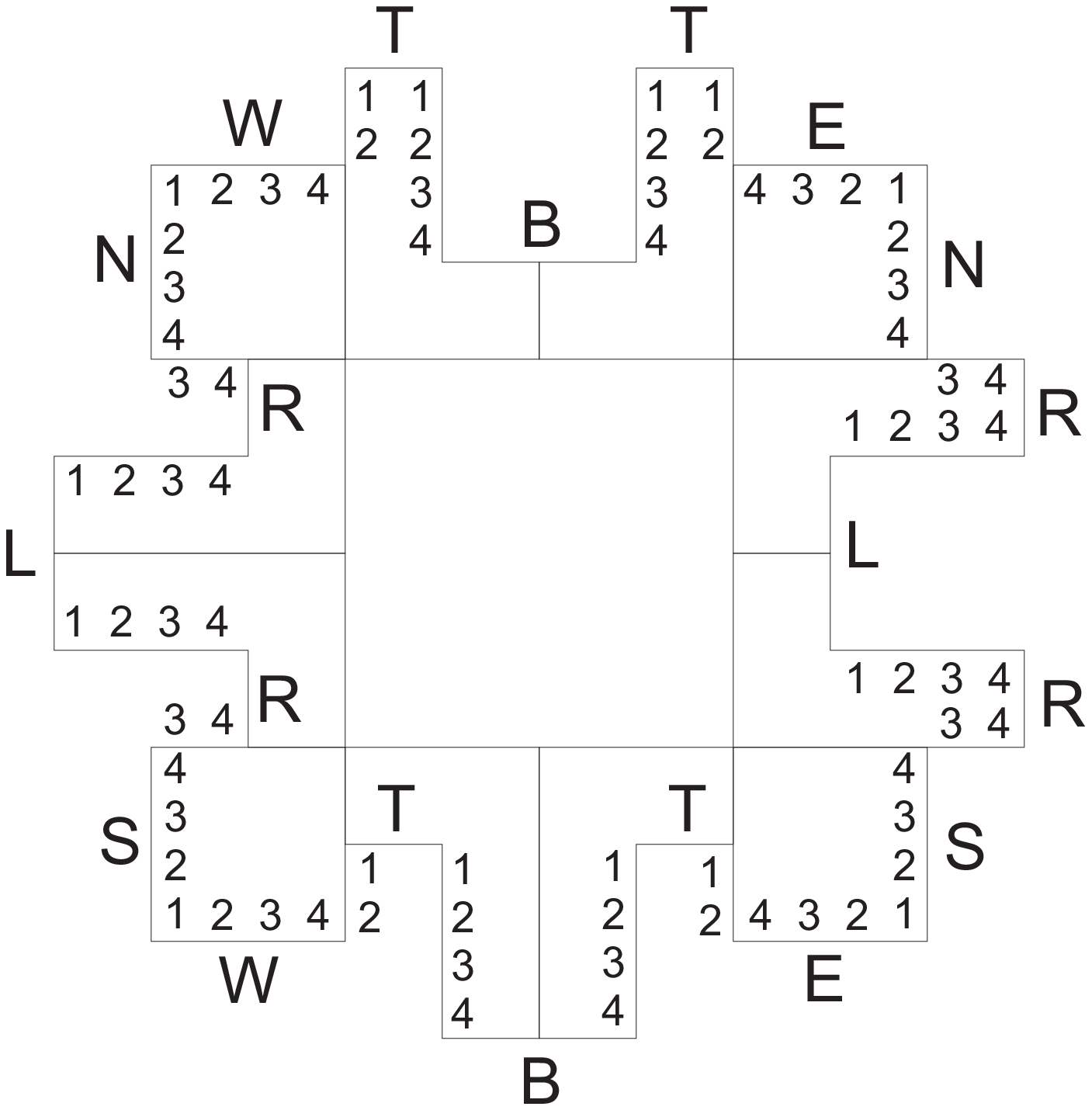}
\caption{\label{fig:full-connectivity-labels} \small The logical depiction of a sample block, showing the labels exposed by the glues on each edge (which are similar for all blocks).  They are all strength-$1$, and the glues on the corners each have a letter and a number, while glues on the binary teeth have only one or the other.}
\end{figure}

\begin{figure}
\centering
\includegraphics[width=6.0in]{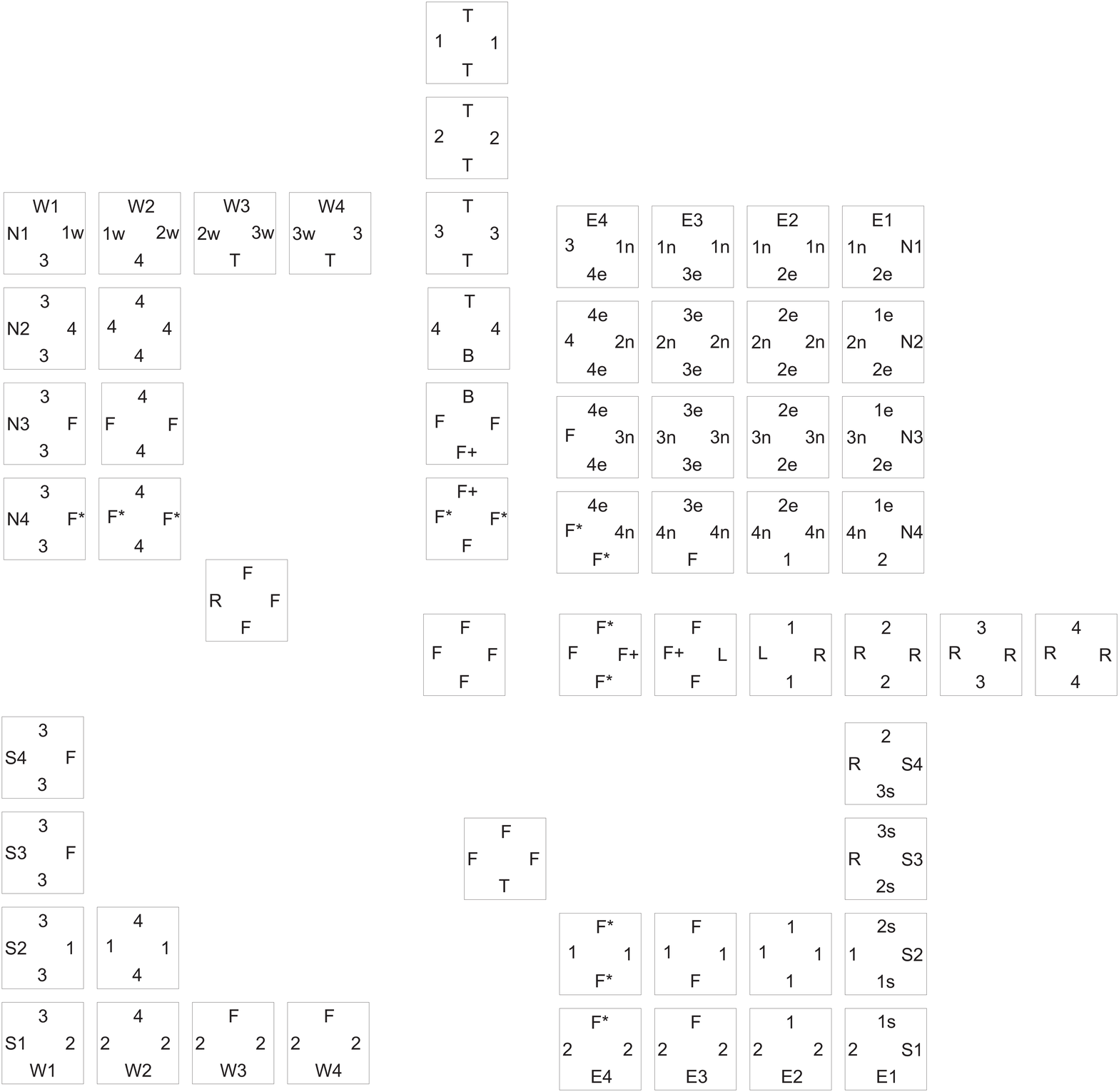}
\caption{\label{fig:full-connectivity-tiles} \small The DNA tile types for the tiles which form the DNA blocks in the fully connected construction.}
\end{figure}

The actual cast assembly begins with the red path of tiles.  It then proceeds in a clockwise direction, as a single tile-wide path which makes a full cycle around the eventual location of the tiles forming the block.  As this path of tiles forms each edge, cooperation between the tiles forming the path and those containing the information specific to that edge (the colored rows) allows the correct pattern of bumps and dents to be formed, as well as allowing the correct RNA tiles to present the necessary glues for attachment by DNA tiles.  Note that all tiles forming the cast (the path and the colored bars along the edges) are RNA tile types.  Every position inside the cast is filled by a DNA tile type.  The glues of the RNA tiles forming the cast interact with the glues of the DNA tiles on, and only on, every north or east edge of an RNA tile on the inside-most row of the cast (which is the boundary between the RNA and DNA tiles).  Since the resulting construction is fully-connected, all tile edges on the outside of a DNA tile block must have glues that ultimately bind to the glues of any neighboring DNA blocks in the final assembly.  This means that locations in the RNA cast without positive strength glue interactions with the DNA tile block have glue mismatches during the block formation (namely, the null glue on the side of the RNA tile and a strength-$1$ glue on the abutting side of a DNA tile). The pattern of glue labels on the outside edges of the blocks of DNA tiles is shown in Figure \ref{fig:full-connectivity-labels}.  The tile set for the DNA tiles making up the blocks in this construction is shown in Figure \ref{fig:full-connectivity-tiles}.

\begin{figure}
\centering
\includegraphics[width=6.5in]{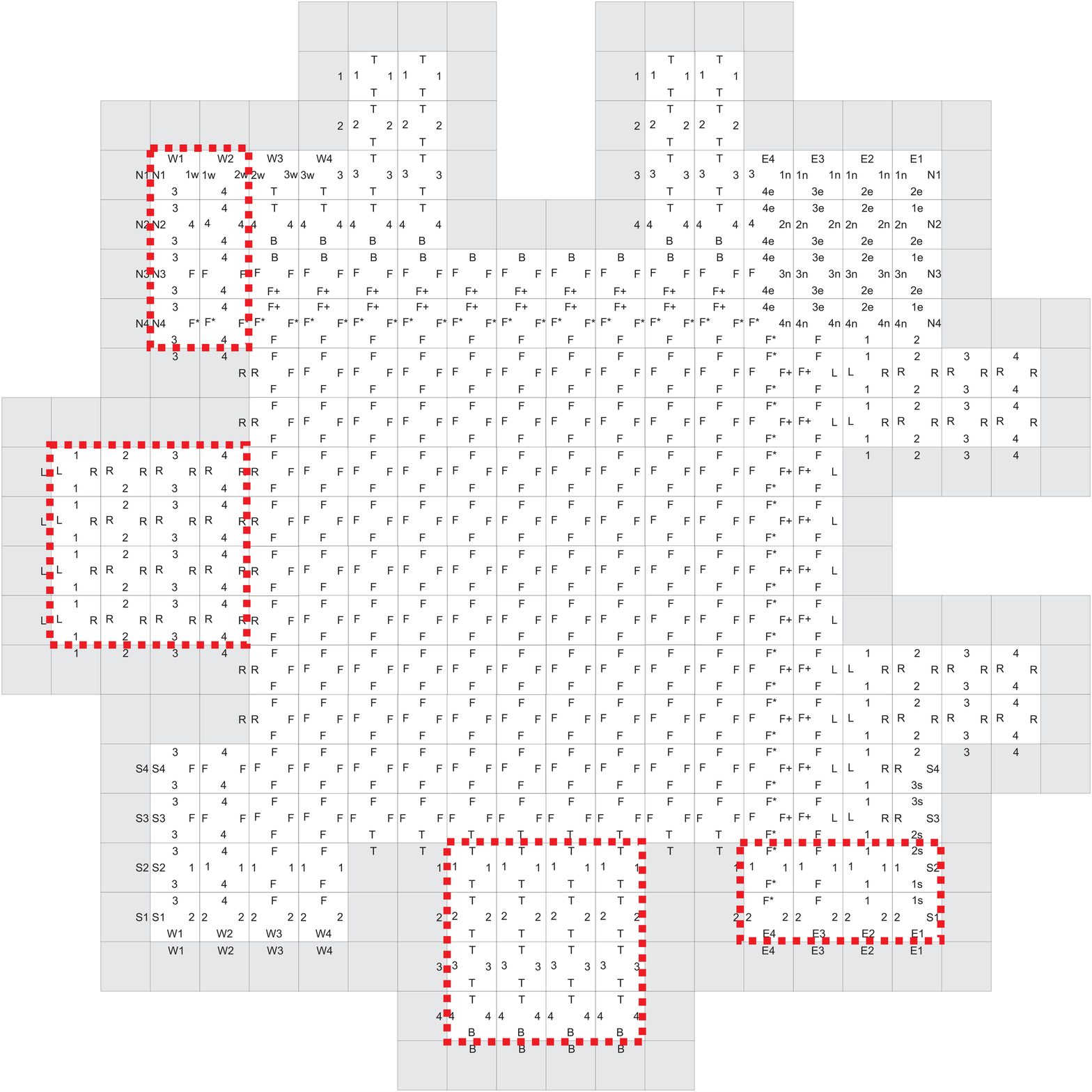}
\caption{\label{fig:full-connectivity-example} \small An example cast and filled in block in the fully connected construction.}
\end{figure}

An example of a block of DNA tiles (pictured in white) along with the innermost row of RNA tiles forming the cast (pictured in grey) is shown in Figure \ref{fig:full-connectivity-example}.  Recall that all DNA tiles have strength-$1$ glues on every edge.  For the RNA tiles in the figure, only the glue labels which interact with positive strength with DNA tiles are shown, and all such labels are shown.  Note that the pattern enforces that every DNA tile which attaches must do so with exactly two input sides --- its south and west sides.

The shape of the path that forms the cast, along with the pattern of non-negative glues exposed and the direction of growth, ensures that the majority of the tiles forming the DNA block cannot attach until the entire cast is complete.  The only exceptions are the tiles surrounded by red boxes in the figure, which can assemble before the cast is completely finished forming.

The resulting blocks of DNA tiles are fully connected internally and have strength-$1$ glues on every exposed edge.  The patterns of glue labels are designed so that complementary edges of separate blocks will bind by fully connecting at every abutting tile edge, but can only connect with positive strength if they are perfectly aligned and completely interlocked and thus encoding the same binary number in their binary teeth.

\subsection{Details of the Full Addressability Construction}
\label{sec-full-addr-details}

\textbf{Unpacking Process:} In the unpacking phase of the current (``full addressability'') construction, the description of the target shape $X$ is decompressed from an algorithmically compact description in a similar manner as it is in the previous two constructions (see Figure~\ref{fig:pod-construction-overview} for a high-level schematic of this process). However, the Turing machine that performs the unpacking algorithm must take into account the following cases: a particular side of some rectangular supertile might have no connection interfaces, the entire side might be a connection interface (as is the case for the previous $O(\log n)$-scale factor constructions), or one side might be the host of several---albeit a finite number of---connection interfaces.

\textbf{Self-Assembly of the RNA Cast: } Similar to the construction for Theorem~\ref{pod_theorem_fully_connected}, we use a cast (or ``mold'') of RNA tiles that assembles an outline of each rectangular supertile. The reason for doing this is to maintain full connectivity of the unique terminal assembly. However, unlike in the construction for Theorem~\ref{pod_theorem_fully_connected}, the self-assembly of the cast in the current full-addressability construction must propagate all the addressability information (e.g., a full specification of which tiles should be labeled ``1'' and which tiles should be labeled ``0'') for the interior of the rectangular supertile \emph{as well as} maintain consistency between all of the glue labels on the connection interfaces of abutting rectangular supertiles. Throughout the discussion of the self-assembly of the RNA cast, we will refer to Figure~\ref{fig:fully-addressed-and-connected-rect}.

The \emph{first} main difference between the (self-assembly of the) RNA cast for the current construction and that of the construction for Theorem~\ref{pod_theorem_fully_connected} is that the former must propagate the addressability information \emph{into} each rectangular supertile. The \emph{second} main difference is that, because of the first main difference, i.e., the cast must propagate information \emph{into} a supertile, the external labels of all of the east-facing glues (for example) on a supertile must match the corresponding west-facing glues (for example) on the adjacent-to-the-east supertile. In fact, great care must be taken to ensure that all of the external glues on all four sides of a rectangular supertile match the external glues on the abutting side of any adjacent rectangular supertiles.

In other words, each side of a supertile must not only potentially accept (addressability) information as input, but it must also ``know'' (but not necessarily use) the addressability information of the opposite side of an abutting rectangular supertile. The RNA cast in the current construction passes addressability information into the west, south and east sides of a supertile. We do not pass addressability information into a supertile via its north side (for a technical reason that we will discuss below) and therefore the (north-facing glues along the) north side of each supertile only needs to ``know'' about the south side of the adjacent-to-the-north supertile. The west side of each supertile needs to ``know'' about the east side of the adjacent-to-the-west supertile as well as the south side of the adjacent-to-the-north supertile (so that this information can be propagated up to the north side of the supertile via its interior). The south side of each supertile does not need to ``know'' about any side of any supertile. Finally, the east side of each supertile must ``know'' about the west side of any adjacent-to-the-east supertile as well as the south side of any adjacent-to-the-north-east supertile. Encoding all of this ``knowledge'' into each of the sides of each of the supertiles in the construction only results in a constant (independent of the target shape $X$) size blowup in the overall tile complexity of our construction.

Note that all tiles in Figure~\ref{fig:fully-addressed-and-connected-rect} that are not shaded solid cannot attach until the completion of the self-assembly of the RNA cast (the solid tiles can attach in the presence of a partially-assembled cast). Once the cast is complete (excluding the four-tile-long linear gadget for the top right corner, which is described in more detail later), the outlined blue tile directly above the red tile in the lower left corner of the supertile (see Figure~\ref{fig:fully-addressed-and-connected-rect}) binds and initiates the bulk of the self-assembly of the rectangular supertile can proceed.

\begin{figure}[htp]
\centering
\includegraphics[width=4.0in]{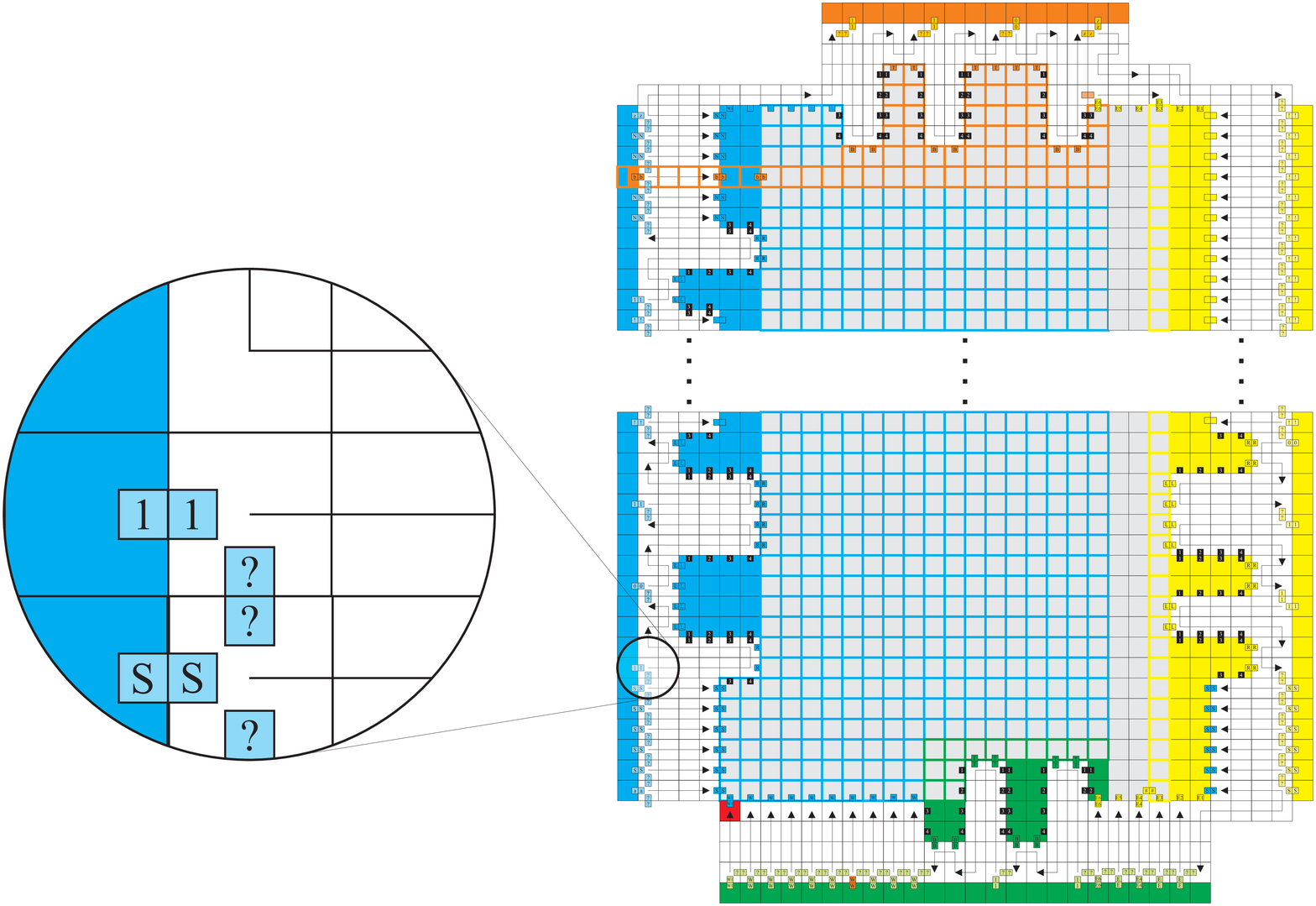} \caption{\label{fig:good_stuff_1} \small The ``?'' glue represents the cast asking the question: ``how should I form the next binary tooth?'' The ``1'' glue tells the RNA cast how to form the next binary tooth in the connection interface of the supertile. Hidden (among many other details) is the length $k$ binary string that encodes the to-be-propagated addressability information for the subsequent four rows in the supertile that are outlined in---or shaded with---blue. Note that if the east/west glue happened to be ``!'' then the cast would simply not form a bump or a dent in the supertile.}
\vspace{-10pt}
\end{figure}

\begin{figure}[htp]
\centering
\includegraphics[width=4.0in]{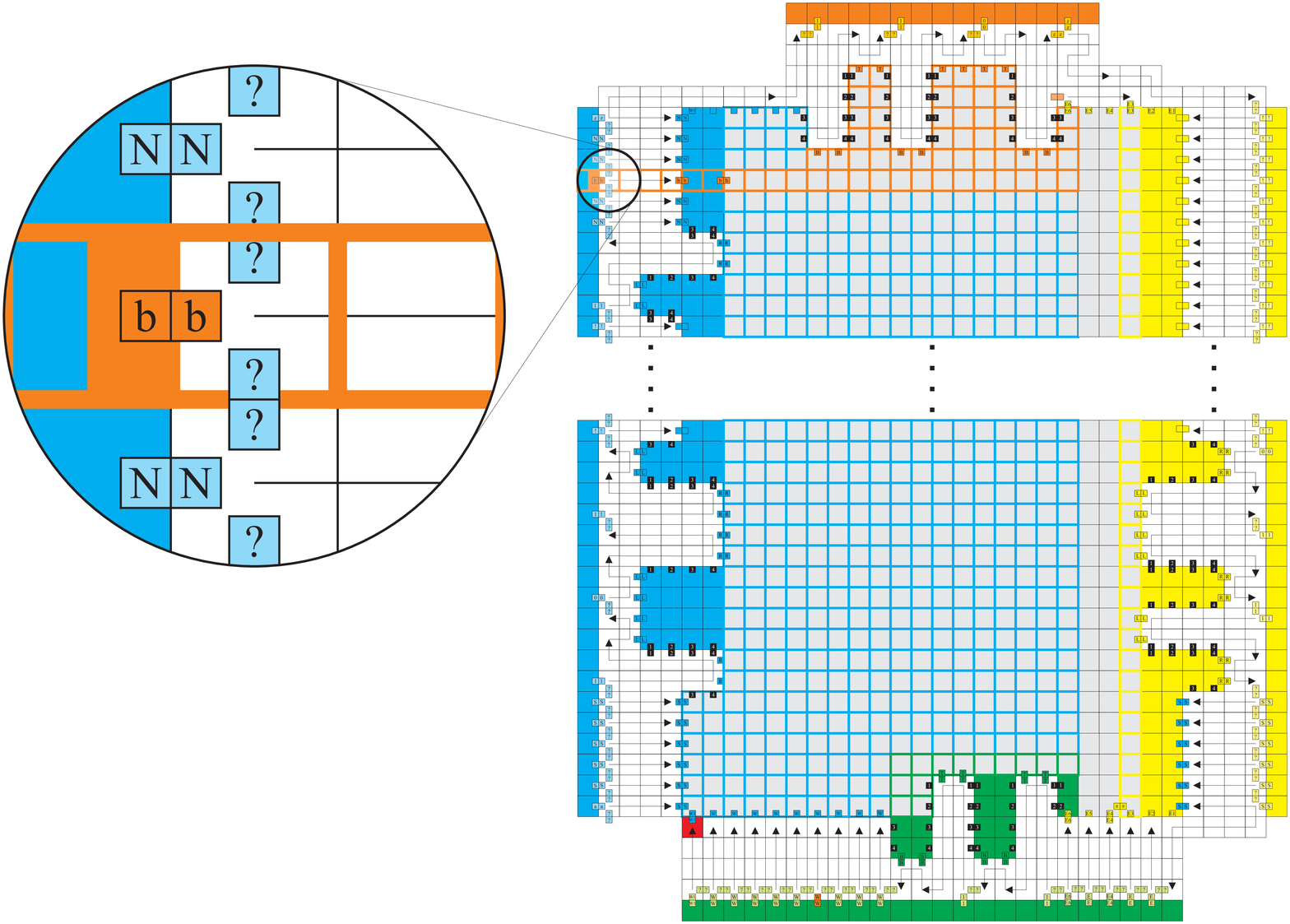} \caption{\label{fig:good_stuff_2} \small In this case, the ``?'' glue represents the cast asking the question: ``How should I proceed?'' The east/west ``b''-labeled orange glue ``responds'' with: ``you're almost at the top of the supertile so start propagating the addressability information for all of the orange outlined tiles eastward.'' All of the addressability information for the orange outlined tiles is encoded into the ``b'' glue label, propagated through a single strength glue to the cast and then ultimately into the interior of the supertile.}
\vspace{-10pt}
\end{figure}

\textbf{Self-Assembly of the Rectangular Supertiles:}
Since the width of rectangular supertiles for this construction is assumed to be bounded by $k$, the glues that bind the tiles of the RNA cast to the outer-most tiles of the rectangular supertile can essentially encode binary strings of length $k$, i.e., the addressability information for each row of the supertile. The addressability information is stored in the outer most solid bars (except for the orange bar) in Figure~\ref{fig:fully-addressed-and-connected-rect}, extracted by the cast via periodic ``cooperation points'' (see Figure~\ref{fig:good_stuff_1}) and propagated through the cast and into every row of the supertile.
In order to propagate this addressability information to each column to the left of the sixth-column-from-the-right, we force all tiles in the rectangular supertile to the left of the sixth-column-from-the-right to initially bind via \emph{only} their south and west sides (this feature is similar to the construction for Theorem~\ref{pod_theorem_fully_connected}). Doing so essentially allows us to assume that every tile type involved in the self-assembly of a rectangular supertile belongs to one of exactly $k$ logical groups of tile types (one for each column). Note that, because of this south/west binding constraint, we do not propagate addressability information of the orange outlined tiles in from the north (in fact, allowing orange outlined tiles to be able to initially bind via their north and west sides while allowing other tiles to initially bind via their west and south sides would result in non-deterministic binding in the fourth-from-the-top row of orange outlined tiles). Instead, we propagate the addressability information for all of the orange outlined tiles via the west side of the supertile through a single strength (orange) bond (see Figure~\ref{fig:good_stuff_2}).

In contrast to the tiles to the left of the sixth-column-from-the-right, we force all of the tiles to the right of---and including---the fifth-to-the-right column (the yellow outlined column of tiles in Figure~\ref{fig:fully-addressed-and-connected-rect}) to bind via their north and east sides. The reason for doing this is because we propagate their addressability information \emph{to the left} through the east-facing glues along east side of the supertile. We purposely avoid propagating the addressability information for the seven right-most columns of the supertile in from the west in order to avoid the problem of:
\begin{quote}
``the east side of a supertile $R$ having to ``know'' about the west side of an adjacent-to-the-east supertile $S$ whose west side must ``know'' about the east side of itself along with the west side of an adjacent-to-the-east supertile $T$  (so the east side of $S$ and the west side of $T$ are consistent) meaning that the east side of $R$ must ``know'' about the east \emph{and} west sides of $S$ \emph{along with} now the west side of $T$...(now imagine what the east side of $R$ must ``know'' if $U$ is an adjacent-to-the-east supertile of $T$)...''
\end{quote}

\begin{figure}[htp]
\centering
\includegraphics[width=4.5in]{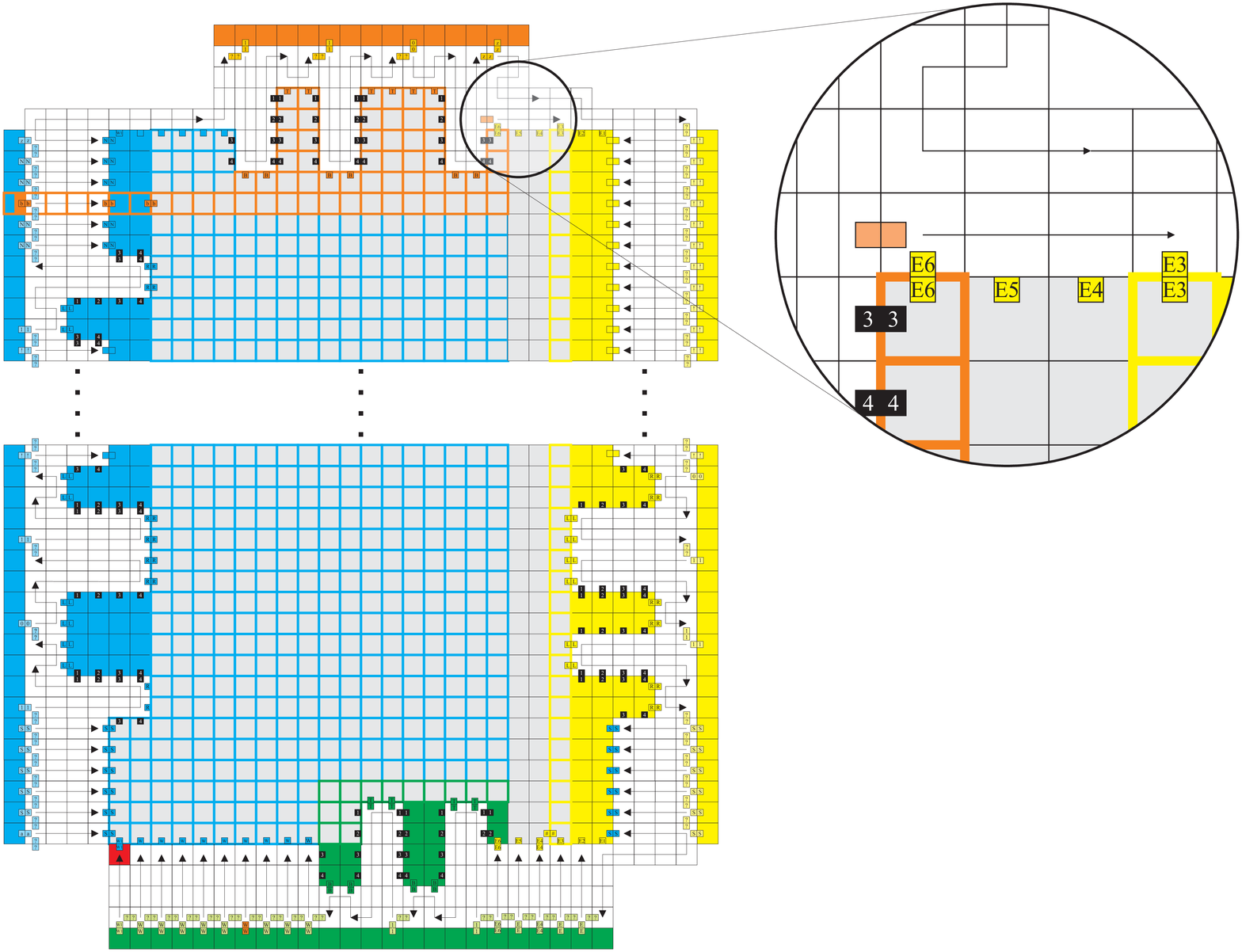} \caption{\label{fig:good_stuff_3} \small The four-tile-long linear gadget, connected via double strength bonds (represented by the right-pointing arrow; the arrow could just as well point to the left as we are assuming a two-handed self-assembly model).  }
\vspace{-10pt}
\end{figure}

\begin{figure}[htp]
\centering
\includegraphics[width=4.5in]{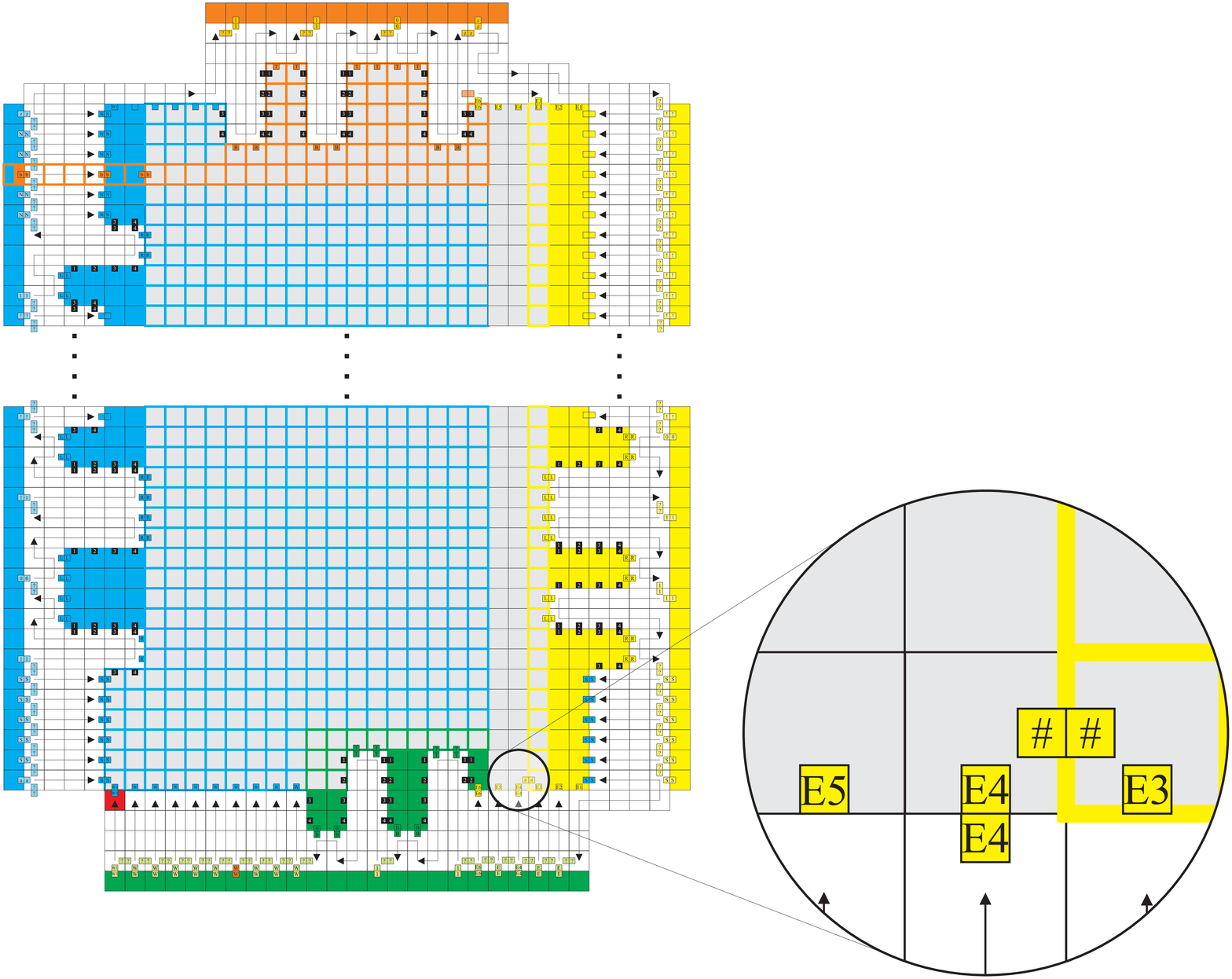} \caption{\label{fig:good_stuff_4} \small The initiation of the self-assembly of the seventh- and sixth-from-the-right columns in the interior of the supertile. This final assembly process is initiated by the cooperation of the bottommost outlined yellow tile cooperates with its adjacent-to-the-south-west neighbor (on the cast). }
\vspace{-10pt}
\end{figure}

Eventually the topmost outlined orange tile in the sixth-from-the-right column (in Figure~\ref{fig:fully-addressed-and-connected-rect}) will bind and subsequently allow the four-tile-long linear gadget (see Figure~\ref{fig:good_stuff_3}) to attach in a two-handed fashion to the upper right corner of the supertile.
The south glue of the rightmost tile in this gadget initiates the southward-growing assembly of the fifth-from-the-right column of outlined yellow tiles (in Figure~\ref{fig:fully-addressed-and-connected-rect}). The final (bottommost) tile in this column cooperates with the RNA cast (see Figure~\ref{fig:good_stuff_4}) to initiate the self-assembly of the seventh- and sixth-from-the-right columns of tiles to self-assemble---these are the final two columns of the supertile to be filled in.

\begin{figure}[htp]
\centering
\includegraphics[width=4.0in]{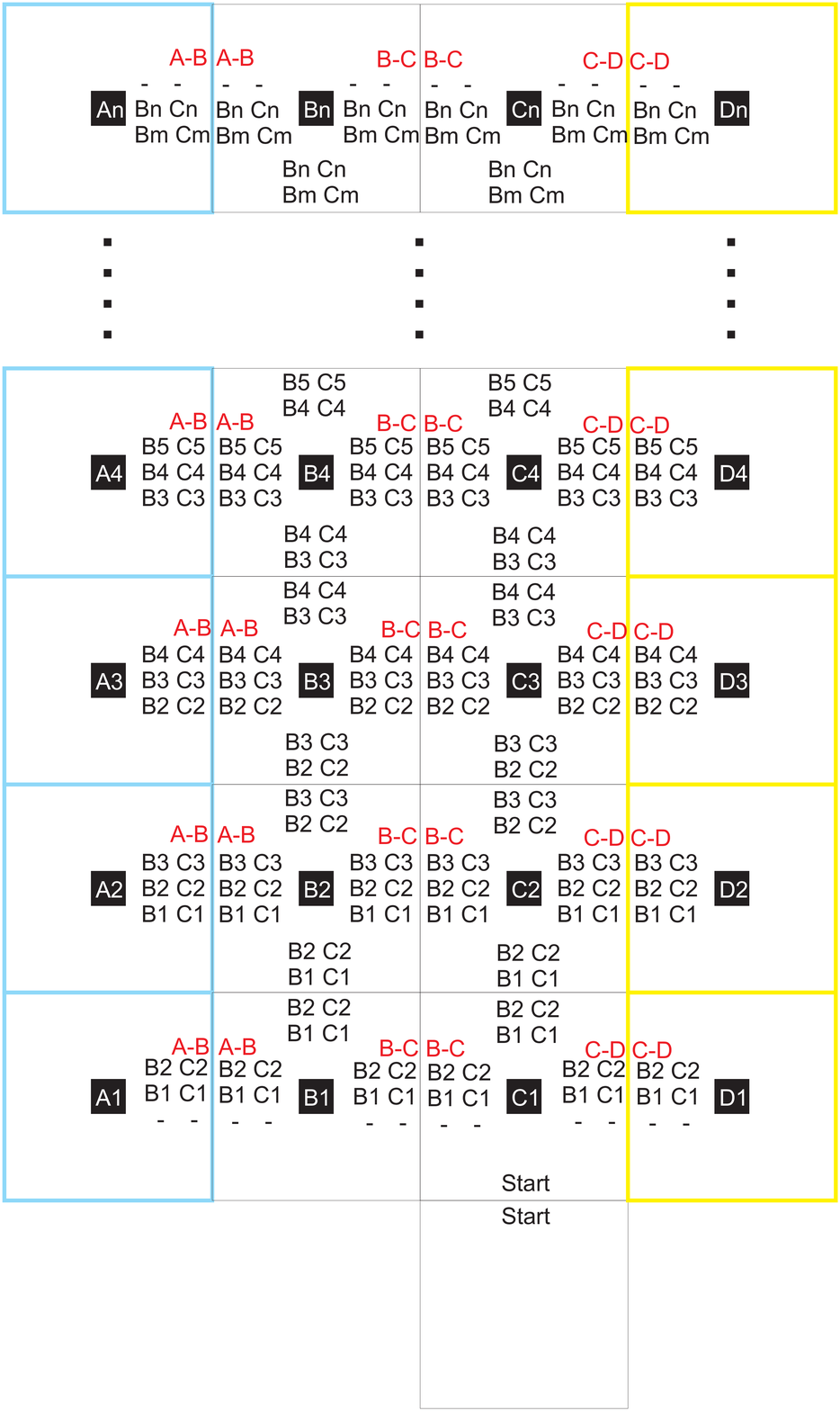} \caption{\label{fig:final-columns} \small Details of formation of the final two columns to form in each block (represented by the `B' and `C' columns).  Let $n$ be the total number of rows in the block and $m = n - 1$.  The two red strings `A-B' and `B-C' are literals.  The values A1-A5, B1-B5, C1-C5, and D1-D5 are variables whose values can be either $0$ or $1$. East and west facing glues are composed of one literal and $6$ variables while north and west facing glues are composed of $4$ variables.}
\vspace{-10pt}
\end{figure}

Note that, in general, the tiles in seventh- and sixth-from-the-right columns can bind initially via any combination of at least two input sides (this is because, at this point, the eighth- and fifth-from-the-right columns are entirely filled in). Thus, care must be taken in order to correctly propagate/maintain the addressability information of these final two columns from \emph{both} the west and the east sides and to handle the nondeterministic ordering in which these columns can form. This can be seen in Figure ~\ref{fig:final-columns}.

\textbf{Self-Assembly of the Target Shape:}
Once the initial stage becomes terminal, and therefore all of the rectangular supertiles which compose the rectangle decomposition of $S$ have completely self-assembled, a $\BREAK$ stage occurs.  After all $RNA$ tiles have dissolved, the rectangular supertiles are free to self-assemble $S$.  This completes the construction.

\end{document}